\documentclass[article]{elsarticle}
\makeatletter
\def\ps@pprintTitle{%
	\let\@oddhead\@empty
	\let\@evenhead\@empty
	\def\@oddfoot{\centerline{\thepage}}%
	\let\@evenfoot\@oddfoot}
\makeatother

\usepackage{lineno,hyperref}
\usepackage{threeparttable}
\usepackage{float}
\usepackage{booktabs}
\usepackage{amsmath}
\usepackage{comment}
\usepackage{color}

\usepackage{amsfonts}
\usepackage{amssymb}
\usepackage{amsmath}
\usepackage{amsthm}
\usepackage{amssymb}

\usepackage{epstopdf}
\usepackage{natbib}

\modulolinenumbers[5]

\begin{document}

\begin{frontmatter}

\title{Atomistic and mean-field estimates of effective stiffness tensor of nanocrystalline materials of hexagonal symmetry}

\author[]{Katarzyna Kowalczyk-Gajewska\corref{author}}
\author[]{Marcin Ma\'zdziarz}

\cortext[author] {Corresponding author.\\\textit{E-mail address:} kkowalcz@ippt.pan.pl}
\address{Institute of Fundamental Technological Research Polish Academy of Sciences, Warsaw, Poland}

\begin{abstract}
Anisotropic core-shell model of a nano-grained polycrystal is extended to estimate the effective elastic stiffness of several metals of hexagonal crystal lattice symmetry. In the approach the bulk nanocrystalline material is described as a two-phase medium with different properties for a grain boundary zone and a grain core. While the grain core is anisotropic, the boundary zone is isotropic and has a thickness defined by the \emph{cutoff radius} of a corresponding atomistic potential for the considered metal.  The predictions of the proposed mean-field model are verified with respect to simulations performed with the use of the Large-scale Atomic/Molecular Massively Parallel Simulator, the Embedded Atom Model, and the molecular statics method. The effect of the grain size on the overall elastic moduli of nanocrystalline material with random distribution of orientations is analysed.
\end{abstract}

\begin{keyword}
 Molecular statics\sep
 Elasticity\sep
 Polycrystal\sep
 Effective medium\sep
 Hexagonal symmetry
\end{keyword}

\end{frontmatter}

\section{Introduction}
\label{sec:Int}

In nanocrystalline materials, usually defined as those polcyrystalline media for which the average grain size is less than 100\,nm \cite{Gleiter00,Gao13}, a significant number of atoms occupies the grain boundary zone or the grain boundary affected zone \cite{Barai11}. Therefore such materials can be treated as composed of two main phases. The effect of grain boundaries on the effective properties of a bulk nanocrystalline material is the more pronounced the smaller is a grain size \cite{Sanders97,Gao13}. An impact of the atom arrangements at the nanoscale on the effective properties of such materials has been studied mainly by means of atomistic simulations \cite{Hanh15}, although some experimental data, in majority related to fcc materials, can be also found in the literature, e.g. \cite{Legros00,Sharma03}. Much less investigations were performed for hexagonal crystals in spite of the technological importance of magnesium, titanium or cobalt.
 
At the macro-level the continuum mechanics description is applicable for nanocrystalline materials, so the mean-field estimates are employed to describe their bulk properties. An extensive review of such estimates was performed by \cite{Kowalczyk18}, in view of which for nano-grained polycrystals the following types of two-phase or multi-phase frameworks were formulated: i.
simplified mixture rule-based models \cite{Carsley95,Kim99,Benson01,Qing06,Zhou07}, ii. inclusion-matrix models \cite{Sharma03} or iii. composite sphere / generalized self-consistent-type models \cite{Jiang04,Chen06,Capolungo07,Mercier07, Ramtani09,Sevostianov2006,SEVOSTIANOV20071304,Kowalczyk18}.

{In the present study on nanocrystalline hexagonal metals we follow the proposal of \cite{Kowalczyk18} for fcc copper, which is inspired by earlier works by \cite{Jiang04} and \cite{Capolungo07}. This composite grain model is formulated in two variants called \emph{the Mori-Tanaka (MT) and self-consistent (SC) core-shell model}, respectively. In view of the proposed geometrical idealization of nanocrystalline medium an additional phase that forms an uniform \emph{isotropic} coating around the \emph{anisotropic} grain core is introduced. Let us mention that a more sophisticated treatment of a grain boundary zone can be found in \cite{Sevostianov2006,SEVOSTIANOV20071304} -- studies dedicated to metal-matrix composite reinforced by nanosized inclusions. Following \cite{Shen03}, authors assumed that the interphase layer between the inclusion and the matrix has isotropic properties which {vary smoothly} with "upward convexity". Alternatively, a step-wise gradation of interphase properties has been assumed by \cite{Jurczak_2019}.  As demonstrated in \cite{Kowalczyk19} also in the frame of the core-shell model inhomogeneous shell properties can be assumed, though, on the cost of a more complicated formulation and necessity to identify additional material parameters.}

{Most often to identify those parameters and validate the proposed estimates the molecular dynamics/statics simulations are used \cite{Schiotz99,Chang03,Choi12,Gao13,Mortazavi2014,Fang16,Kowalczyk18}. Finite element calculations are scarce because they require a non-standard constitutive models accounting for size effects \cite{Kim20123942}. The common trend observed in the majority of such simulations is reduction of elastic stiffness with a decreasing grain size \cite{Zhao06,Gao13,Xu17}. Such variation of elastic moduli with a grain size would be predicted by the core-shell models when, on average, the boundary zone is elastically less stiff than a grain core \cite{Jiang04,Ramtani09,Gao13,Kowalczyk18}. It is worth mentioning that a reverse trend was found in atomistic simulation by \cite{Kowalczyk19} for two (i.e. vanadium and niobium) out of eight metals of cubic symmetry studied therein. Interestingly, these two crystals have a Zener anisotropy factor lower than one, contrary to remaining six metals.}

The challenging issue for those multi-phase concepts is to propose an appropriate description of a grain boundary zone (or zones), namely its volume fraction, morphology and local properties. To this end, likewise, molecular static/dynamic simulations are employed, commonly in a bi-crystal configuration, e.g. \cite{Kluge1990,Singh18}. Results depend on the disorientation axis and angle between two grains, see also \cite{Rittner96,Tschopp07}. For a mean-field model of random nanocrystalline medium the average properties representative for all types of boundaries are of interest, therefore we apply the procedure adopted in \cite{Kowalczyk19}. Elastic properties of a grain boundary zone are identified on generated polycrystal samples for which the fraction of transient shell atoms encompasses the whole volume. A thickness of this zone is assumed as equal to the \emph{cutoff radius} of a respective atomistic potential.
	
{The present paper reports a follow-up to the recent studies by \cite{Kowalczyk18,Kowalczyk19}.
The goal of this research is to evaluate  applicability of the core-shell model proposed therein for describing the effective elastic stiffness of nanocrystalline metals of hexagonal lattice symmetry. In particular, the assumptions concerning the description of a grain boundary zone are verified.}

{The paper is constructed as follows.  The successive section presents details of spectral decomposition of elasticity tensor for crystals of hexagonal symmetry, which due to the properties of the fourth order tensors is equivalent to a transverse isotropy case. The possible anisotropy measures for such tensor are also discussed. Moreover, this section reminds the formulation of a core-shell model and shows how its different variants can be obtained from the general formula.  Section \ref{sec:Cm} is devoted to fundamentals of atomistic simulations. Comparison of the results of atomistic simulations and  core-shell model predictions is performed in Section~\ref{sec:Res} (detailed results of molecular simulations are collected in \ref{App}.). The last section contains summary and conclusions.}

\section{Two-phase core-shell model for bulk nanocrystals of hexagonal symmetry}
\label{sec:Cont}

The anisotropic linear law between the stress 
$\boldsymbol{\sigma}$ and strain 
$\boldsymbol{\varepsilon}$ in the grain is assumed, namely
\begin{equation}\label{locconst}
\boldsymbol{\sigma}=\mathbb{C}(\phi_c)\cdot\boldsymbol{\varepsilon},\quad
\boldsymbol{\varepsilon}=\mathbb{S}(\phi_c)\cdot\boldsymbol{\sigma},\quad\mathbb{S}(\phi_c)\mathbb{C}(\phi_c)=\mathbb{I}\,,
\end{equation}
where $\mathbb{C}(\phi_c)$ and $\mathbb{S}(\phi_c)$ are the fourth order elastic stiffness and compliance tensors of a given symmetry.  Argument $\phi_c$ denotes symbolically an orientation of local axes
$\{\mathbf{a}_k\}$ with respect to some macroscopic frame
$\{\mathbf{i}_k\}$. $\mathbb{I}$ is a fourth order symmetrized identity tensor. 

A unit cell of crystal lattice with a hexagonal closed packed (hcp) spatial distribution of atoms has a six-fold rotational symmetry axis $\mathbf{c}$. Therefore for hcp crystals the local elastic stiffness tensor  $\mathbb{C}(\phi_c)$ exhibits transverse isotropy. It means that, from the point of view of hcp unit cell geometry \cite{Kocks00}, as concerns elastic properties only orientation of $\mathbf{c}$ axis matters, while orientation of $\mathbf{a}_i$ axes (e.g. the so-called armchair or zigzag one) does not influence the form of $\mathbb{C}(\phi_c)$. The spectral form of the fourth order tensor of transverse isotropy is \cite{Walpole81,Rychlewski95,Kowalczyk09}
\begin{equation}\label{Eq:SpekTrans}
\mathbb{C}(\phi^c)=h_1\mathbb{P}_1^{ti}(\xi,\phi^c)+h_2\mathbb{P}_2^{ti}(\xi,\phi^c)+2G_2\mathbb{P}_3^{ti}(\phi^c)+2G_3\mathbb{P}^{ti}_4(\phi^c)\,,
\end{equation}
where $\mathbb{P}_i^{ti}$ are fourth order orthogonal projectors of the form
\begin{eqnarray}
\label{Eq:P1P2}
\mathbb{P}_1^{ti}(\xi,\phi^c)+\mathbb{P}_2^{ti}(\xi,\phi^c)&=&\mathbb{I}^{\rm{P}}+\frac{1}{6}(3\mathbf{N}-\mathbf{I})\otimes(3\mathbf{N}-\mathbf{I})\,,\\
\label{Eq:P3}
\mathbb{P}_3^{ti}(\phi^c)&=&\frac{1}{2}\left(\left[(\mathbf{I}-\mathbf{N})\otimes(\mathbf{I}-\mathbf{N})\right]^{T(23)+T(24)}\!-\!
(\mathbf{I}-\mathbf{N})\otimes(\mathbf{I}-\mathbf{N})\right)\,,\\ \label{Eq:P4}
\mathbb{P}_4^{ti}(\phi^c)&=&\frac{1}{2}
\left[\mathbf{N}\otimes(\mathbf{I}-\mathbf{N})+(\mathbf{I}-\mathbf{N})\otimes\mathbf{N}\right]^{T(23)+T(24)}
\end{eqnarray}
with $(\mathbb{A}^{T(23)+T(24)})_{ijkl}\equiv(\mathbb{A})_{ikjl}\!+\!(\mathbb{A})_{ilkj}$ and $\mathbf{N}(\phi^c)=\bar{\mathbf{c}}(\phi^c)\otimes\bar{\mathbf{c}}(\phi^c)$. Unit vector $\bar{\mathbf{c}}$ is a normalized axis of a hcp unit cell: $\mathbf{c}/|\mathbf{c}|$.
\begin{figure}
	\centering
	\includegraphics[width=0.85\textwidth]{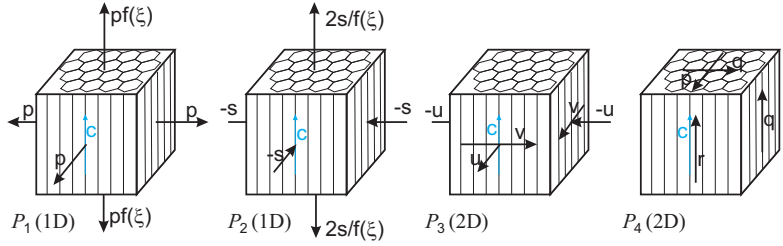}
		\caption{Illustration of eigen-subspaces of the elasticity tensor of hexagonal symmetry}
	\label{fig:TranswSub}
\end{figure}
Two single Kelvin moduli $h_1$ and $h_2$ are two single eigenvalues of the $2\times 2$ matrix 
\begin{equation}\label{Eq:L2x2}
\left[\begin{array}{cc}
3K& L_{12}\\
L_{12}& 2G_1\end{array}\right]
\end{equation}
where:
\begin{eqnarray}
3K&=&(2C_{1111}+C_{3333}+2C_{1122}+4C_{1133})/3,\\ 2G_1&=&(C_{1111}+2C_{3333}+C_{1122}-4C_{1133})/3,\\ 
L_{12}&=&\sqrt{2}(C_{3333}-C_{1111}+C_{1133}-C_{1122})/3\,,
\end{eqnarray}
while in-plane $G_2$ and out-of-plane $G_3$ shear moduli are specified as:
\begin{equation}
G_2=(C_{1111}-C_{1122})/2\,,\quad G_3=C_{1313}
\end{equation}
$C_{ijkl}$ are the components of the elasticity tensor $\mathbb{C}$ in the orthonormal basis for which $\mathbf{i}_3=\mathbf{c}$. 

Four strictly positive Kelvin moduli: $h_K$ ($K=1,2$), $2G_2$ and $2G_3$ correspond to four eigen-subspaces of strain or stress states established by the elasticity tensor, which are respectively:
\begin{itemize}
	\item two one-dimensional subspaces of axially symmetric stretching along $\mathbf{c}$. The specification of these two subspaces depends on the value of stiffness distributor $\xi$ (more details can be found in \ref{Ap:1}),
	\item the two-dimensional subspace of in-plane pure shears (i.e. pure shears in the isotropy plane which is a plane perpendicular to $\mathbf{c}$ axis),
	\item the two-dimensional subspace of out-of-plane pure shears (i.e pure shears in the plane containing $\mathbf{c}$ axis).
\end{itemize}
This subspaces are schematically illustrated in Fig. \ref{fig:TranswSub}. For the states belonging to the respective subspaces the proportionality is observed between stress and strain tensors. It should be mentioned that if $L_{12}$ equals zero then the space $P_1$ is the space of hydrostatic states, $P_2$ the space of isochoric axially symmetric stretching and $h_1=3K$, $h_2=2G_1$. 

As discussed by \cite{Kowalczyk19} in the case of cubic crystal the Zener anisotropy factor $\zeta$ enables the assessment of an anisotropy degree but also distinction between anisotropy types. Cubic crystal is elastically anisotropic if $\zeta\neq 1$ and crystals can be classified as those for which $\zeta<1$ and those for which $\zeta>1$. In the case of hexagonal (transverse isotropic) crystals definition of a unique parameter of such property is not possible. Instead, a set of three parameters is proposed, which play a similar role as the Zener parameter, namely:
\begin{equation}\label{Eq:zeta1}
\zeta=\{L_{12},G_2/G_1,G_3/G_1\}\,.
\end{equation}
Hexagonal crystal is in fact isotropic if and only if $\zeta=\{0,1,1\}$. Six subclasses of transverse isotropy may be distinguished depending if the ratios $\zeta^{II}=G_2/G_1$, $\zeta^{III}=G_3/G_1$ are larger or smaller than 1 (note that they are always positive) and on their relative value so if $\zeta^{II}>\zeta^{III}$ or reversely. A subclass of materials for which $L_{12}=0$ is called volumetrically isotropic. Note that for such materials hydrostatic state is an eigenstate in the spectral decomposition (\ref{Eq:SpekTrans}), similarly to the case of isotropic material. It should be stressed that anisotropy degree as such can be also assessed using a single scalar, for example the universal anisotropy factor \cite{OstojaStarzewski} or the anisotropy measure $\zeta_0$ (Eq. \ref{Eq:zeta2}), which is based on the closest isotropic approximation. However, two latter anisotropy factors do not enable us to distinguish between transverse isotropy subclasses.     
 
The standard micromechanical theories treat coarse-grained polycrystals as one-phase heterogeneous materials. In the elastic regime heterogeneity of strain and stress fields results from the varying orientation of crystal axis $\mathbf{c}$ in the polycrystalline representative volume element (RVE). Estimates of effective response of the hcp grain aggregate are obtained on the basis of knowledge of the local elastic properties and the assumed  micro-macro transition scheme. The formulas for the standard estimates, such as the Voigt, Reuss, Hashin-Shtrikman or self-consistent one, can be found in \ref{Ap:1}. These estimates are not sensitive to the grain size. A fundamental difference as compared to cubic polycrystals studied within similar framework by \cite{Kowalczyk18,Kowalczyk19} is that, as long as $L_{12}\neq 0$, the overall bulk modulus for random polycrystal is different from the local one and varies between the schemes.

{As discussed in the Introduction, for nanocrystalline materials the common way to assess the effective properties of the bulk material is to use a two-phase model. In the present research the core-shell model developed in \cite{Kowalczyk18} is used with different properties for a grain boundary zone and a grain core. While the grain core is anisotropic, the boundary zone surrounding the core is isotropic.  The model enables estimation of the effective stiffness tensor $\bar{\mathbb{C}}$ for an arbitrary orientation distribution. By fundamental theories of micromechanics \cite{Sevostianov18} such tensor relates the averaged strain  $\mathbf{E}=\langle \boldsymbol{\varepsilon}\rangle $ and stress $\boldsymbol{\Sigma}=\langle \boldsymbol{\sigma}\rangle $ in the polycrystalline RVE, namely: 
	\begin{equation}
		\boldsymbol{\Sigma}=\bar{\mathbb{C}}\cdot\mathbf{E}\,
	\end{equation}
where $\langle . \rangle =\frac{1}{V}\int_V(.) dV $ denotes averaging performed over the representative material volume.

An idea behind the core-shell model is to calculate effective stiffness by exploiting the double inclusion scheme of \cite{HoriNematNasser93}. Accordingly the coated grain is embedded in the infinite medium of the stiffness $\mathbb{C}_{\rm{m}}$ taken equal to  $\mathbb{C}_{\rm{s}}$ or $\bar{\mathbb{C}}_{\rm{CS}}$ for Mori-Tanaka (MT) or self-consistent (SC) variants of the model, respectively. As a result it is obtained
\begin{equation}\label{eq:core-shell}
\bar{\mathbb{C}}_{\rm{CS}}=\left[f_0\mathbb{C}_{\rm{s}}\mathbb{A}_{\rm{s}}+(1-f_0)\left<\mathbb{C}(\phi^c)\mathbb{A}(\phi^c)\right>_{\mathcal{O}}\right]\left[f_0\mathbb{A}_{\rm{s}}+(1-f_0)\left<\mathbb{A}(\phi^c)\right>_{\mathcal{O}}\right]^{-1}
\end{equation}					
where
\begin{equation}
\mathbb{A}(\phi^c)=(\mathbb{C}(\phi^c)+\mathbb{C}_*(\mathbb{C}_{\rm{m}}))^{-1}(\mathbb{C}_{\rm{m}}+\mathbb{C}_*(\mathbb{C}_{\rm{m}}))\,,
\end{equation}	
\begin{equation}
\mathbb{A}_{\rm{s}}=(\mathbb{C}_{\rm{s}}+\mathbb{C}_*(\mathbb{C}_{\rm{m}}))^{-1}(\mathbb{C}_{\rm{m}}+\mathbb{C}_*(\mathbb{C}_{\rm{m}}))
\end{equation}
and  $\mathbb{C}_*(\mathbb{C}_{\rm{m}})$ is the Hill tensor \cite{Hill65}. Quantity $f_0$ is the volume fraction of the grain boundary zone.  It is calculated by the formula
\begin{equation}\label{def:fsa}
f_0=1-\left(1-\frac{2\Delta}{d}\right)^3\,,
\end{equation} 
where $d$ is an averaged grain diameter and $\Delta$ -- the coating thickness. The formula is found assuming the spherical shape of grain cores and the coating. Previous studies indicated \cite{Kowalczyk18,Kowalczyk19} that $\Delta$ can be assumed as equal to the \emph{cutoff radius} of the atomistic potential valid for the considered metal. Presence of the ratio ${2\Delta}/{d}$ makes the estimate $\bar{\mathbb{C}}_{\rm{CS}}$ sensitive to the grain size. More details on the model formulation can be found in the mentioned papers.}  
The isotropic shell properties need to be identified separately. In the present work, following \cite{Kowalczyk19}, they are established by means of atomistic simulations by analyzing polycrystalline aggregates with a very small grains, in which the grain boundary zone encompasses whole grains. Note that Eq. (\ref{eq:core-shell}) can be understood in a generalized fashion enabling one to encompass also another two-phase schemes applicable to nanocrystalline media known in the literature. For example, a simple mixture rule-based model (Voigt's iso-strain scheme) is obtained assuming $\mathbb{A}(\phi^c)=\mathbb{A}_s=\mathbb{I}$, while Reuss' iso-stress scheme is recovered when $\mathbb{A}(\phi^c)=\mathbb{C}(\phi_c)^{-1}$ and $\mathbb{A}_s=\mathbb{C}_s^{-1}$.
 
A limit of a coarse-grained polycrystal  is obtained when $f_0\rightarrow 0$, so when the volume fraction of grain boundary zones approaches zero. In such limit the effective properties 
$\bar{\mathbb{C}}_{\rm{CS/SC}}$ approach
the self-consistent estimate of \cite{Kroner58} for a one-phase polycrystal. 
Respective limit estimates of the bulk and shear modulus related to the effective stiffness $\bar{\mathbb{C}}_{\rm{CS/MT}}$, and perfectly random orientation distribution, approach the following values:
\begin{equation}\label{Eq:KsCSMTlim}
\bar{K}_{\rm{CS/MT}}^{\infty}=K-\frac{L_{12}^2}{6G_1+9K_*}
\end{equation}
\begin{equation}\label{Eq:GsCSMTlim}
\bar{G}_{\rm{CS/MT}}^{\infty}=5\left(\frac{1}{G
_1+G_*-\frac{L_{12}^2}{6(K_1+K_*)}}+\frac{2}{G
_2+G_*}+\frac{2}{G
_3+G_*}\right)^{-1}-G_*
\end{equation}
where 
\begin{equation}
K_*=4 G_{\rm{s}}\,,\quad G_*=G_{\rm{s}}\frac{8G_{\rm{s}}+9K_{\rm{s}}}{3(2G_{\rm{s}}+K_{\rm{s}})}
\end{equation}
These values are some lower (resp. upper) bound estimates of $\bar{\mathbb{C}}$ if the difference $\mathbb{C}_{\rm{s}}-\mathbb{C}(\phi_c)$ is negative (resp. positive)  definite for any $\phi_c$. Those bounds lie within less rigorous Reuss and Voigt bounds, which are approached if $G_s$ tends to 0 and $\infty$, respectively. 
Evidently, for another limit value: $f_0\rightarrow 1$ the estimates $\bar{\mathbb{C}}_{\rm{CS/MT}}$ and 
$\bar{\mathbb{C}}_{\rm{CS/SC}}$ are equal and coincide with $\mathbb{C}_{\rm{s}}$, so with the shell properties. 

\section{Computational methods}
\label{sec:Cm}
{The molecular statics (MS) method (i.e. at 0\,K temperature) \cite{Tadmor2011,Maz2010,Maz2011} simulations were performed with the use of the Large-scale Atomic/Molecular Massively Parallel Simulator (LAMMPS) \cite{Plimpton1995}. As an approximation describing the energy between atoms the Embedded Atom Model (EAM) \cite{Tadmor2011,Becker2013277} was used. The Open Visualization Tool OVITO \cite{Ovito2010} was used to analyse and visualize the results of the simulations. 
The methodology for preparing polycrystal samples by the Voronoi tessellation algorithm implemented in the Atomsk program  \cite{Hirel2015212}, their pre-relaxation and atomic simulations was adapted almost straightforwardly from \cite{Kowalczyk18, Kowalczyk19}. All calculation samples were approximately cubes. The size of the samples was chosen so that: \emph{small sample} contained only an amorphous structure representing the grain boundaries, an \emph{medium sample} of about 0.5 million atoms and a \emph{large sample} of about 4 million atoms.
To get the components of stiffness tensor, $\bar{C}_{ijkl}$, for all pre-relaxed structures, the stress-strain method with the maximum strain amplitude of 10$^{-4}$ was utilised \cite{Plimpton1995,Mazdziarz2015}.}
	
\begin{figure}[h!]
	\centering
	\begin{tabular}{cc}
		(a)& (b)\\
	\includegraphics[width=.45\textwidth]{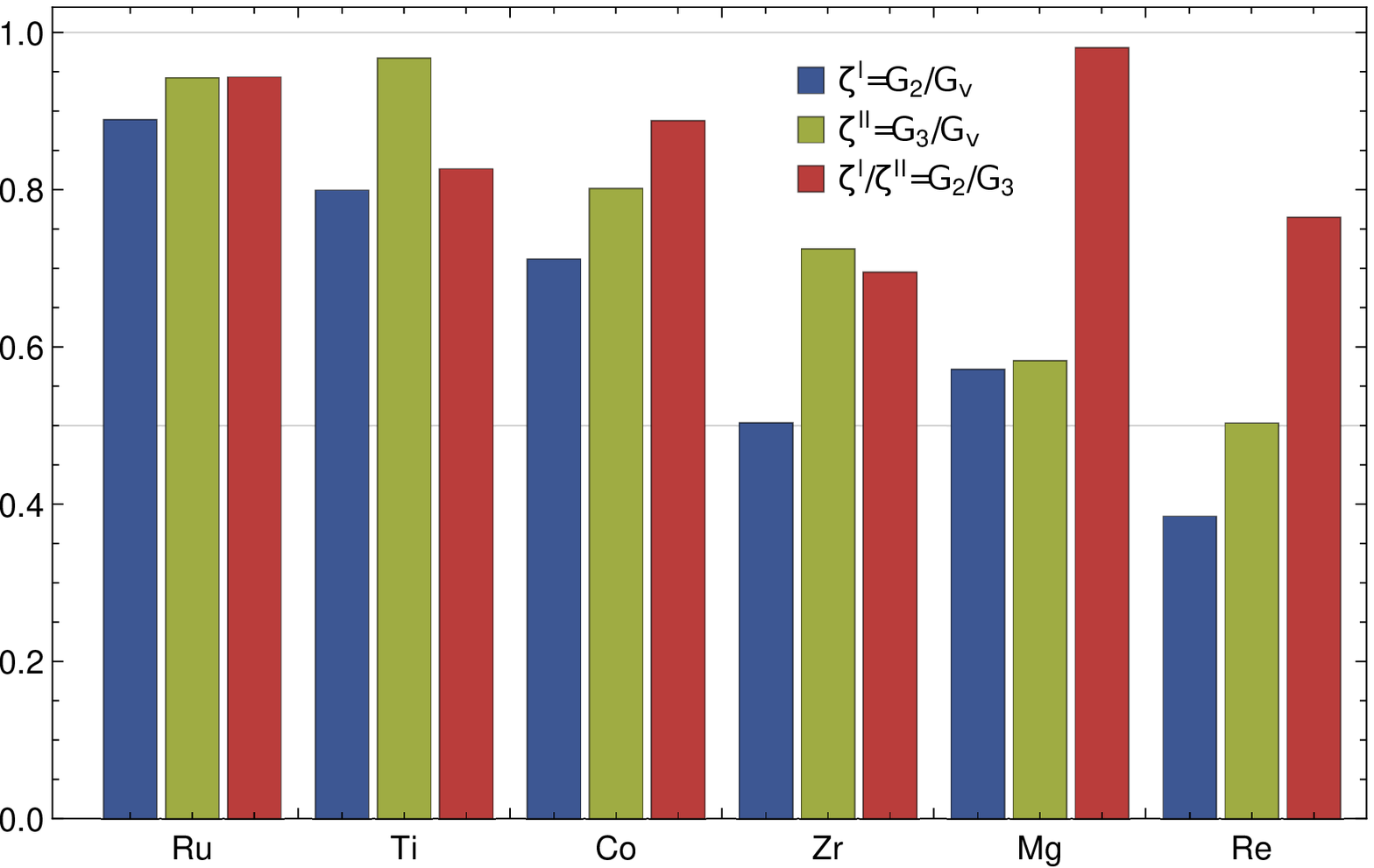}&\includegraphics[width=.45\textwidth]{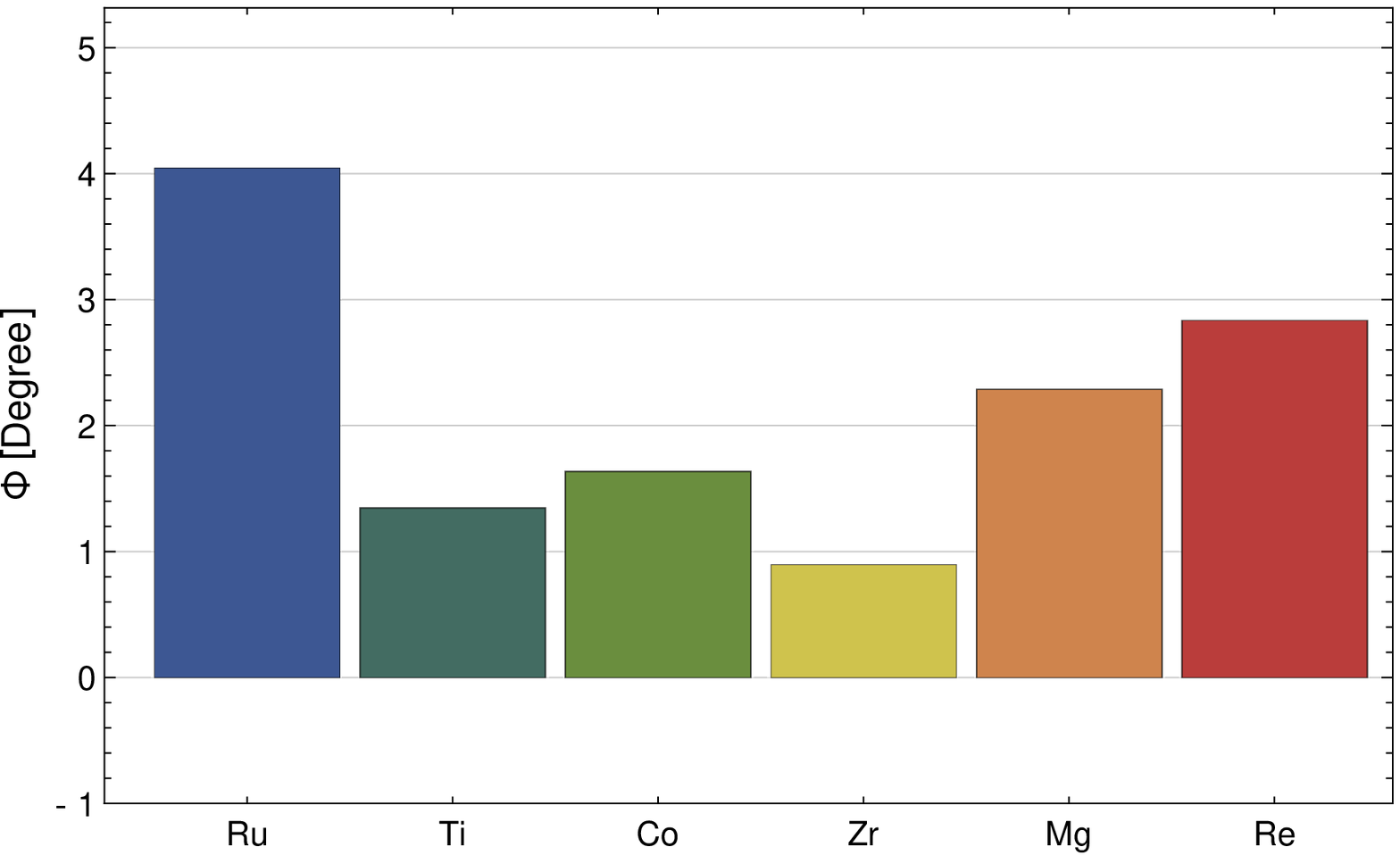}\\
   (c)& (d)\\
   \includegraphics[width=.45\textwidth]{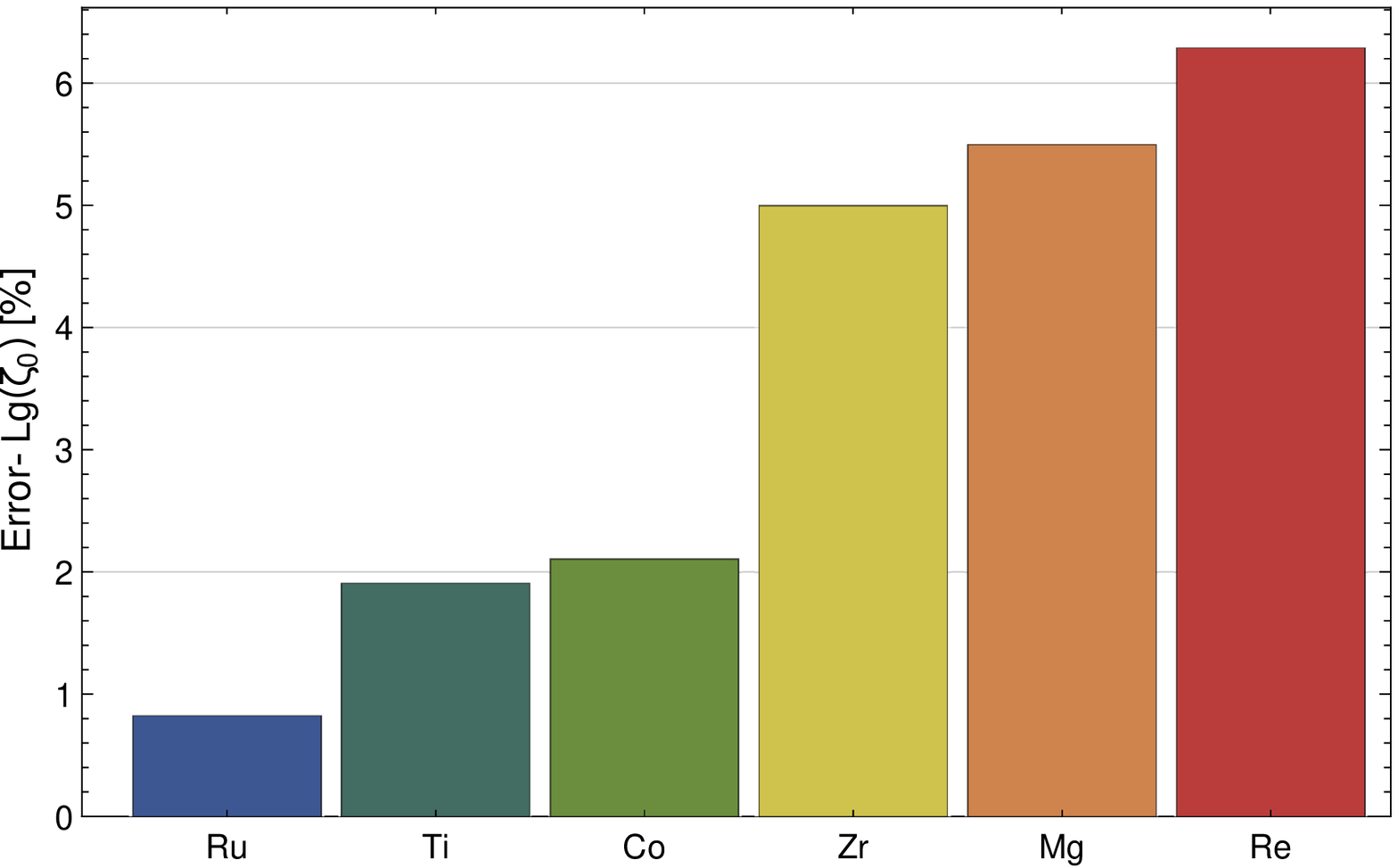}
 &\includegraphics[width=.45\textwidth]{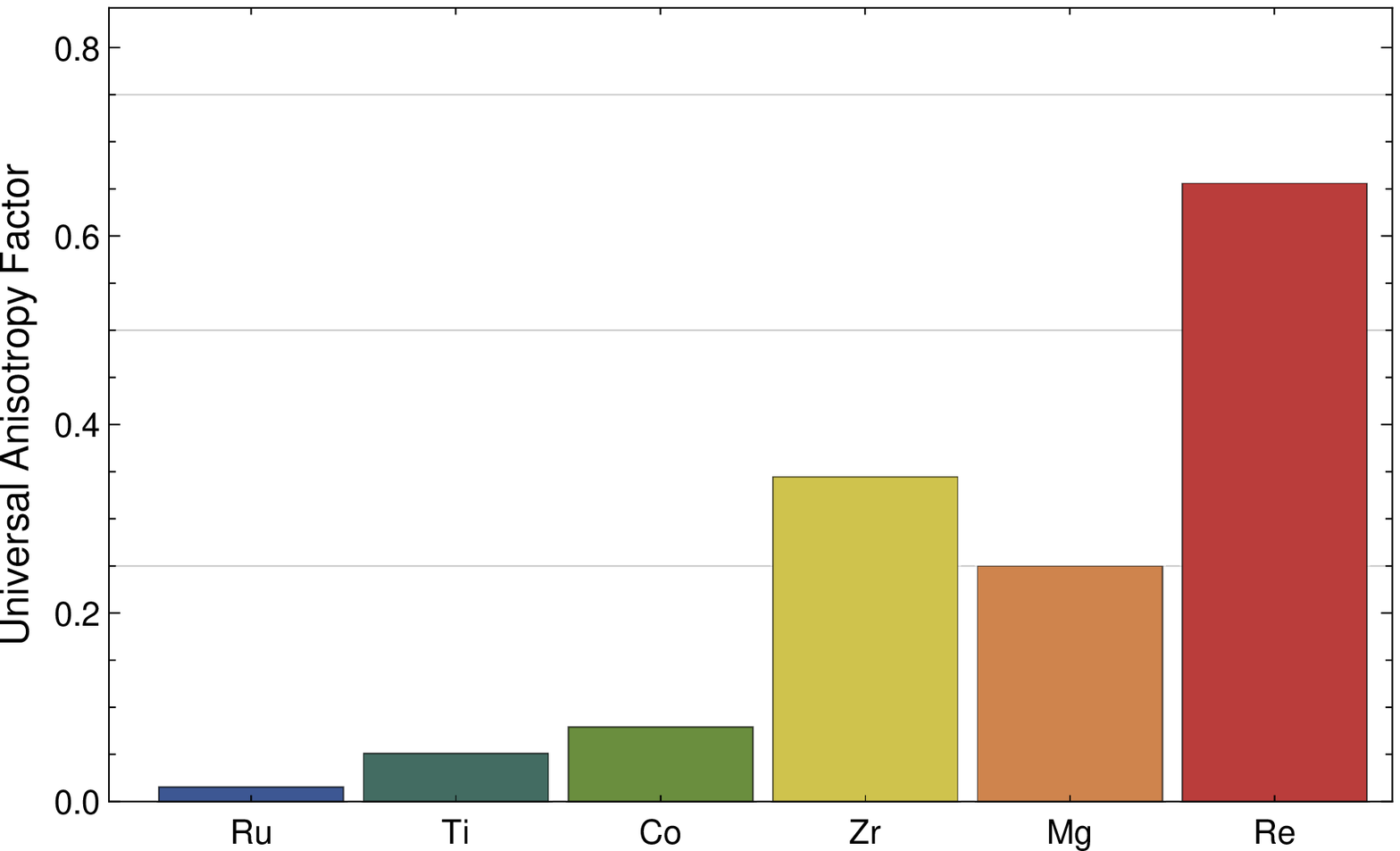}\\
   \end{tabular}
	\caption{Anisotropy measures for considered hcp metals: (a) Zener-like anisotropy factors $\zeta$, (b) Non-caxiality ratio $\Phi$ (see \ref{Ap:1}), (c) universal anisotropy measure $\zeta_0$, (d) universal elastic anisotropy index.}
	\label{fig:Zener}
\end{figure}

In order to study the effect of the anisotropy degree as well as the number and size of grains on mechanical properties of polycrystalline material, six metals of hcp lattice symmetry with seven grain sizes each were considered in this work, see the following enumeration \ref{i:Ru}--\ref{i:Re}, Tabs. \ref{tab:SamplesRu}--\ref{tab:SamplesRe} and Fig. \ref{fig:Zener}.

The stiffness parameters of a grain boundary zone used in the core-shell model should be representative for an averaged stiffness of an interphase layers between any pair of grain orientations. In \cite{Kowalczyk19} it was proposed to identify such parameters by performing atomistic simulations on samples for which the size was reduced so that the fraction $f_0$ of transient shell atoms {approaches unity}. The name of these samples starts with a letter S in Table \ref{tab:SamplesHCP}.

{
\begin{enumerate}[I.]
	\item \textbf{Ruten (Ru)} \label{i:Ru}

The ruten EAM potential parametrized by \cite{Fortini2008} was used. This potential reproduces the hcp-ruten monocrystal equivalent orthogonal cell (but that still respects the hexagonal lattice) lattice constants $a_{hcp}$=2.704\,\AA, $b_{hcp}$=4.684\,\AA, $c_{hcp}$=4.288\,\AA, the cohesive energy $E_{c}$=-6.86\,eV, and the elastic constants in crystallographic axes coinciding with Cartesian coordinate system axes: $C_{1111}$=546.54\,GPa, $C_{3333}$=619.07\,GPa, $C_{1122}$=169.87\,GPa, $C_{1133}$=170.85\,GPa, and  $C_{2323}$=199.58\,GPa. The characteristics of computational ruten samples are listed in the Tab.\ref{tab:SamplesRu}.
	
	\item  \textbf{Titanium (Ti)} \label{i:Ti}
	
The titanium EAM potential parametrized by \cite{Zope2003} was used. This potential reproduces the hcp-titanium monocrystal equivalent orthogonal cell (but that still respects the hexagonal lattice) lattice constants $a_{hcp}$=2.953\,\AA, $b_{hcp}$=5.114\,\AA, $c_{hcp}$=4.681\,\AA, the cohesive energy $E_{c}$=-4.85\,eV, and the elastic constants in crystallographic axes coinciding with Cartesian coordinate system axes: $C_{1111}$=171.47\,GPa, $C_{3333}$=189.96\,GPa, $C_{1122}$=84.23\,GPa, $C_{1133}$=77.07\,GPa, and $C_{2323}$=52.79\,GPa. The characteristics of computational titanium samples are listed in the Tab.\ref{tab:SamplesTi}.

		\item \textbf{Cobalt (Co)} \label{i:Co}
	
The cobalt EAM potential parametrized by \cite{Pun2012} was used. This potential reproduces the hcp-cobalt monocrystal equivalent orthogonal cell (but that still respects the hexagonal lattice) lattice constants $a_{hcp}$=2.519\,\AA, $b_{hcp}$=4.362\,\AA, $c_{hcp}$=4.056\,\AA, the cohesive energy $E_{c}$=-4.39\,eV, and the elastic constants in crystallographic axes coinciding with Cartesian coordinate system axes: $C_{1111}$=310.01\,GPa, $C_{3333}$=357.51\,GPa, $C_{1122}$=145.67\,GPa, $C_{1133}$=119.48\,GPa, and $C_{2323}$=92.54\,GPa. The characteristics of computational cobalt samples are listed in the Tab.\ref{tab:SamplesCo}.
	
	\item \textbf{Zirconium (Zr)} \label{i:Zr}
	
The zirconium EAM potential parametrized by \cite{Mendelev2007} was used. This potential reproduces the hcp-zirconium monocrystal equivalent orthogonal cell (but that still respects the hexagonal lattice) lattice constants $a_{hcp}$=3.230\,\AA, $b_{hcp}$=5.596\,\AA, $c_{hcp}$=5.186\,\AA, the cohesive energy $E_{c}$=-6.02\,eV, and the elastic constants in crystallographic axes coinciding with Cartesian coordinate system axes: $C_{1111}$=174.27\,GPa, $C_{3333}$=211.40\,GPa, $C_{1122}$=109.70\,GPa, $C_{1133}$=80.54\,GPa, and $C_{2323}$=46.45\,GPa. The characteristics of computational zirconium samples are listed in the Tab.\ref{tab:SamplesZr}.
	
	\item \textbf{Magnesium (Mg)} \label{i:Mg}

The magnesium EAM potential parametrized by \cite{Johnson2004} was used. This potential reproduces the hcp-magnesium monocrystal equivalent orthogonal cell (but that still respects the hexagonal lattice) lattice constants $a_{hcp}$=3.199\,\AA, $b_{hcp}$=5.541\,\AA, $c_{hcp}$=5.210\,\AA, the cohesive energy $E_{c}$=-1.55\,eV, and the elastic constants in crystallographic axes coinciding with Cartesian coordinate system axes: $C_{1111}$=55.88\,GPa, $C_{3333}$=69.40\,GPa, $C_{1122}$=28.70\,GPa, $C_{1133}$=20.19\,GPa, and $C_{2323}$=13.86\,GPa. The characteristics of computational magnesium samples are listed in the Tab.\ref{tab:SamplesMg}.
	
		\item \textbf{Rhenium (Re)} \label{i:Re}
	
The rhenium EAM potential parametrized by \cite{Setyawan2018} was used. This potential reproduces the hcp-rhenium monocrystal equivalent orthogonal cell (but that still respects the hexagonal lattice) lattice constants $a_{hcp}$=2.761\,\AA, $b_{hcp}$=4.782\,\AA, $c_{hcp}$=4.477\,\AA, the cohesive energy $E_{c}$=-8.03\,eV, and the elastic constants in crystallographic axes coinciding with Cartesian coordinate system axes: $C_{1111}$=340.24\,GPa, $C_{3333}$=448.68\,GPa, $C_{1122}$=259.96\,GPa, $C_{1133}$=217.92\,GPa, and $C_{2323}$=52.51\,GPa. The characteristics of computational rhenium samples are listed in the Tab.\ref{tab:SamplesRe}.
\end{enumerate}
}

\section{Results}
\label{sec:Res}

\subsection{Results of atomistic simulations}
\label{ssec:ResAS}

The following notation for computational samples of nanocrystalline hcp material subjected to the atomistic simulations is used $$\rm{SIZE}-N_g-\rm{SYS}$$ where $\rm{SIZE}$ is a relative size of sample (S -- small, M -- medium or L -- large) assessed by the number of unit cells in the sample, $N_g$ - a number of orientations of crystal axes (here 16, 54, 125, 128 or 250 randomly selected orientations), while $\rm{SYS}$ denotes the geometry of grain distribution, i.e.: BCC or random, see Tables
 \ref{tab:SamplesRu}--\ref{tab:SamplesRe} presented in \ref{App}.
As in \cite{Kowalczyk18,Kowalczyk19}  orientations are defined in terms of Euler angles. Detailed results, in the form of full elasticity tensors $\bar{\mathbb{C}}$\,, derived from molecular simulations of analysed samples for six hcp metals are collected in the Tables
   \ref{tab:Cij-cRu}, \ref{tab:Cij-cTi}, \ref{tab:Cij-cCo}, \ref{tab:Cij-cZr}, \ref{tab:Cij-cMg} and \ref{tab:Cij-cRe}, respectively. 
Consistently with the previous studies on cubic nanocrystalline metals mentioned above, it is found that the number of orientations and the morphological distribution of grains have much smaller impact on the value of elastic stiffness than a number of atoms per grain, which in the present context is equivalent to the grain size.

\begin{figure}[!h]
	\centering
	\begin{tabular}{ccc}
		\textbf{Rhenium (HCP)}& & \\
		\includegraphics[width=0.33\linewidth]{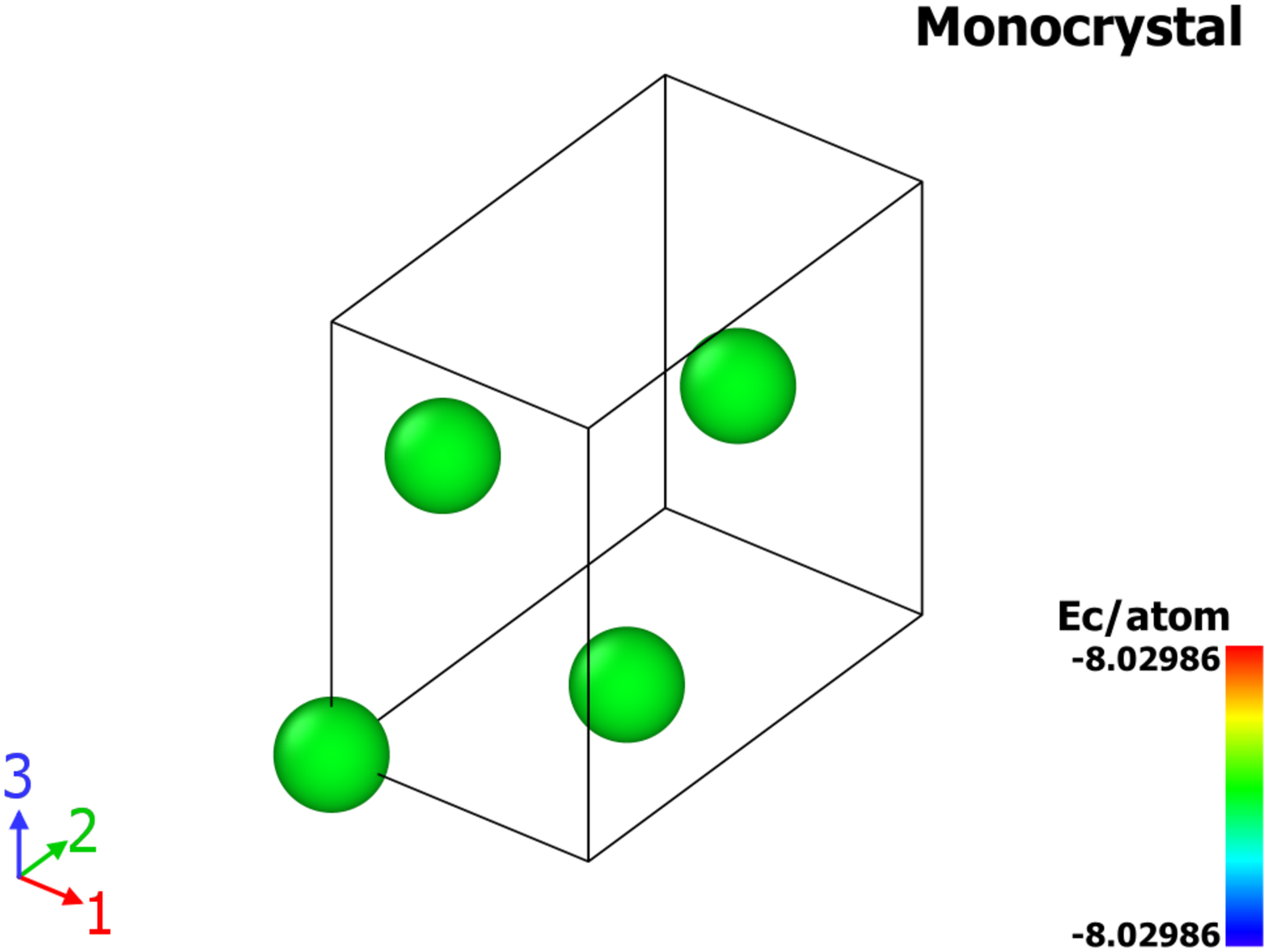} &
		\includegraphics[width=0.33\linewidth]{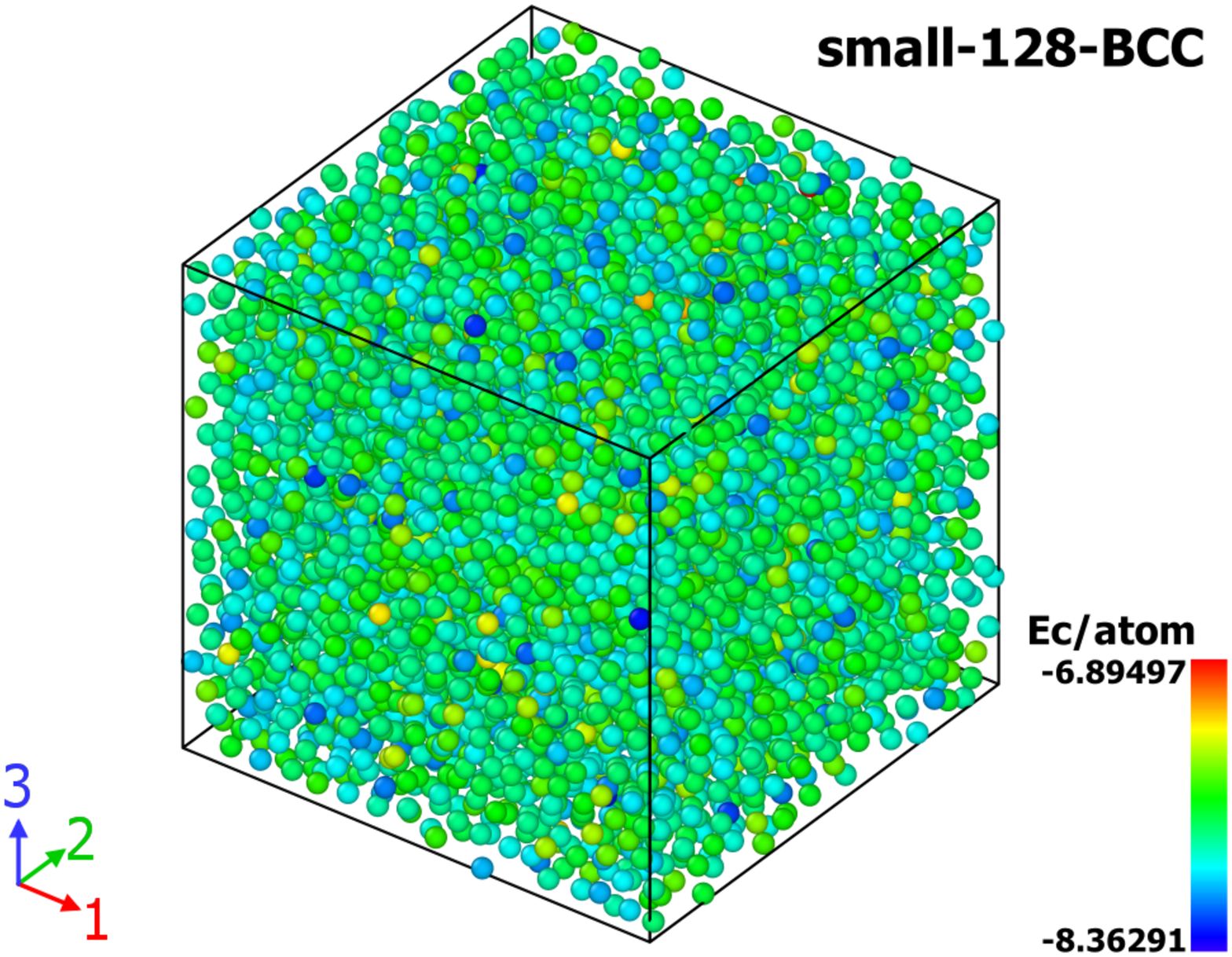} &
		\includegraphics[width=0.33\linewidth]{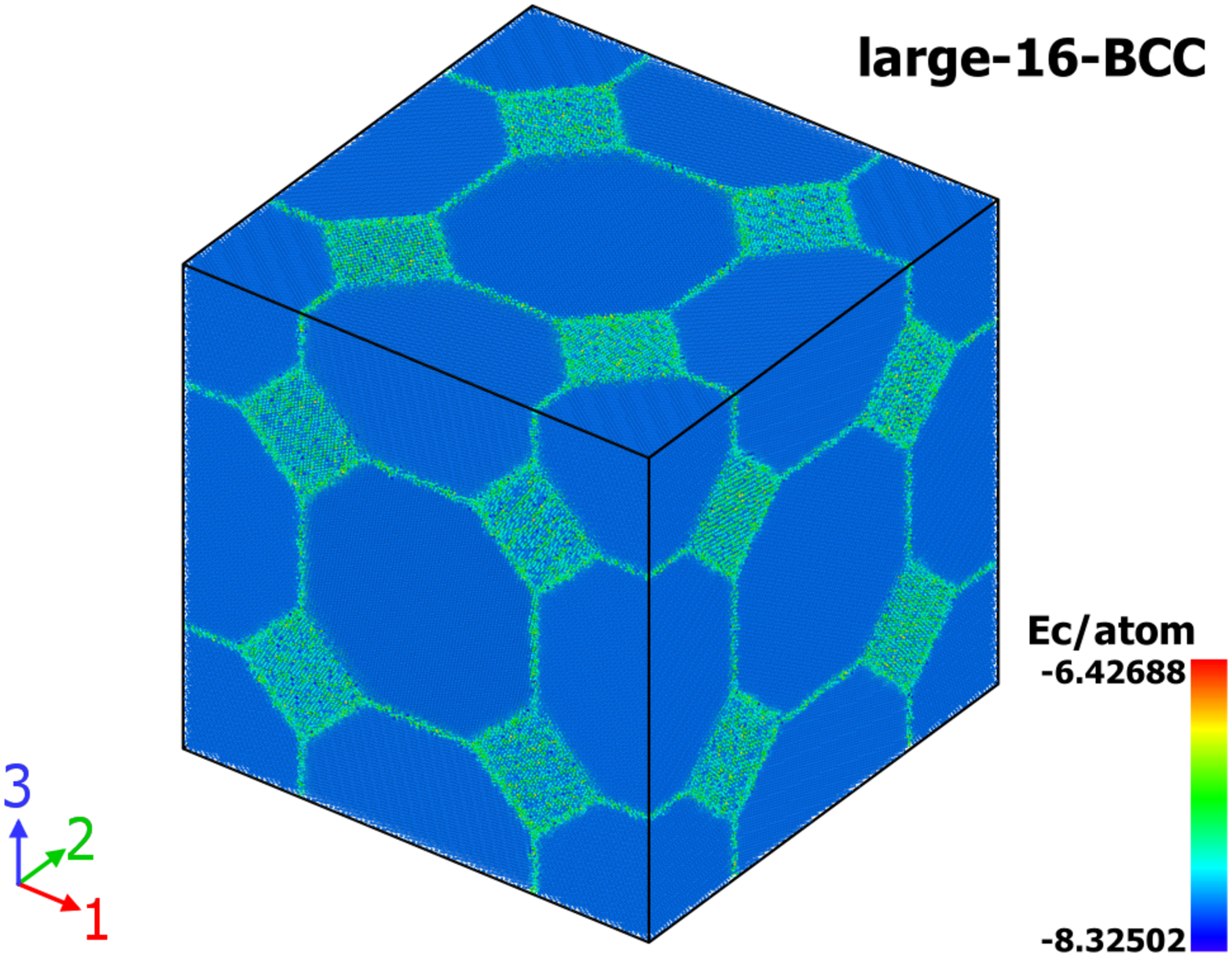} 
	\end{tabular}
	\caption{Visualization of selected atomistic computational samples and cohesive energy Ec (eV/atom) for rhenium.}
	\label{fig:Samples}
\end{figure}

The obvious reason for such correlations can be deduced from Fig. \ref{fig:Samples} where
selected atomistic computational samples and cohesive energy $E_{c}$\,(eV/atom) are visualized. {As it is seen, with a decreasing average grain size, the fraction of transient shell atoms in the sample rises, increasing the impact of the grain boundary zone on the overall response. The samples with a smallest ratio $\rm{SIZE}/N_g$ have almost all atoms belonging to this zone. As $\rm{SIZE}/N_g$ increases samples can be described as a two-phase medium composed of crystalline cores of well-ordered atoms surrounded by amorphous wrapping. It is consistent with the idea of a core-shell model recalled in Sec.\ref{sec:Cont}. Present results indicated that the assumption concerning the assessment of the shell thickness $\Delta$ taken for cubic nanocrystals can be extended to hcp metals, so that $\Delta$ is assumed as equal to \textit{the cutoff radius} of the atomistic potential. The respective values of $d$ and $f_0$ obtained using Eq. (\ref{def:fsa}) are collected in Tables \ref{tab:SamplesRu}, \ref{tab:SamplesTi}, \ref{tab:SamplesCo}, \ref{tab:SamplesZr}, \ref{tab:SamplesMg} and \ref{tab:SamplesRe} in \ref{App}.}
Because the assumed orientation distribution within the samples is random the closest isotropic approximation of the calculated elasticity tensors, collected in Tables \ref{tab:Cij-cRu}-\ref{tab:Cij-cZr} in \ref{App}, is established.

Following previous studies by \cite{Kowalczyk18,Kowalczyk19} the closest isotropic approximation $\bar{\mathbb{C}}^{\mathcal{L}}_{\rm{iso}}$ of anisotropic $\bar{\mathbb{C}}$ is defined employing the Log-Euclidean metric as proposed by \cite{Moakher06}. Using this method two scalars, approximated isotropic bulk modulus $\bar{K}^{\mathcal{L}}_{\rm{iso}}$ and shear $\bar{G}^{\mathcal{L}}_{\rm{iso}}$ are obtained and next compared with the respective estimates found using the core-shell model.
As an universal (i.e. applicable to any material symmetry) anisotropy measures the error $\zeta_0$ resulting from the applied isotropic approximation is used in this work. It is calculated as a normalized difference between $\bar{\mathbb{C}}^{\mathcal{L}}_{\rm{iso}}$,
\begin{equation}\label{Eq:Ciso}
\bar{\mathbb{C}}^{\mathcal{L}}_{\rm{iso}}=3\bar{K}^{\mathcal{L}}_{\rm{iso}}\mathbb{I}^{\rm{P}}+2\bar{G}^{\mathcal{L}}_{\rm{iso}}(\mathbb{I}-\mathbb{I}^{\rm{P}})\,,
\end{equation}
and the actual $\bar{\mathbb{C}}$. It is defined as \cite{Kowalczyk11b}
\begin{equation}\label{Eq:zeta2}
\zeta_0=\frac{||\mathrm{Log}\bar{\mathbb{C}}-\mathrm{Log}\bar{\mathbb{C}}^{\mathcal{L}}_{\rm{iso}}||}{||\mathrm{Log}\bar{\mathbb{C}}||}\times 100\%\geq 0\,,
\end{equation}
where $||\mathbb{A}||=\sqrt{\mathbb{A}\cdot\mathbb{A}}=\sqrt{A_{ijkl}A_{ijkl}}$ and $\mathrm{Log}\mathbb{A}=\sum_{K}\mathrm{log}\lambda_L\mathbb{P}_K$ ($\lambda_K$ - eigenvalues of $\mathbb{A}$, $\mathbb{P}_K$ - eigenprojectors of $\mathbb{A}$ obtained by its spectral decomposition). 
More on the approximation and detailed formulas can be found in \cite{Kowalczyk18,Kowalczyk19}. Another universal anisotropy measures have been discussed in \cite{OstojaStarzewski}. In particular, the non-dimensional quantity defined in terms of Voigt and Reuss estimates of the overall bulk and shear modulus for random polycrystal (see \ref{Ap:1}) has been recommended in that paper. This so-called universal anisotropy index, equal zero for isotropy, is defined as:
\begin{equation}
A^U=5\frac{\bar{G}_{\rm{V}}}{\bar{G}_{\rm{R}}}+\frac{\bar{K}_{\rm{V}}}{\bar{K}_{\rm{R}}}-6\geq 0
\end{equation}

The isotropized bulk and shear moduli, together with the anisotropy factor (\ref{Eq:zeta2}), are collected in Table \ref{tab:SamplesHCP} for the analysed hcp metal samples. For a reader convenience, in the table these samples are ordered according to the increasing averaged grain size.  
For each metal the shell elastic parameters: $K_s$ and $G_s$, were established for the sample with $f_0$ approaching unity, namely ${\rm{S}}-128-{\rm{BCC}}$. The identified values, together with values of $K$, $G_1$, $L_{12}$ and two shear moduli $G_2$ and $G_3$, that is constants defining the Kelvin moduli and stiffness distributor of monocrystals, are collected in Table \ref{tab:Shell-CutOff}. The \emph{cutoff radius} $\Delta$ of the applied atomistic potential is also placed there. Note that this is a set of necessary input data to obtain the predictions of a core-shell model in the next subsection. Metals in this table are ordered according to the increasing value of anisotropy degree measured by $\zeta_0$. For comparison purpose universal anisotropy index is also included.

	\begin{table}[H] 
	\caption{The overall isotropized bulk and shear moduli $\bar{K}^{\mathcal{L}}_{\rm{iso}}$\,[GPa] and $\bar{G}^{\mathcal{L}}_{\rm{iso}}$\,[GPa] and anisotropy measure $\zeta_0$\,[\%] calculated for the effective stiffness tensors resulting from the atomistic simulations for metals of hcp lattice geometry. Samples are ordered according to the increasing average grain size $d$, while metals according to the decreasing anisotropy measure $\zeta_0$ of single crystal (see Table \ref{tab:Shell-CutOff}).}
	\label{tab:SamplesHCP}
	\centering
	\renewcommand{\arraystretch}{1.5}
	\scriptsize 
	\begin{tabular}{|c| c c c| c c c| c c c|}
		\hline
		Sample  & $\bar{K}^{\mathcal{L}}_{\rm{iso}}$ & $\bar{G}^{\mathcal{L}}_{\rm{iso}}$ & $\zeta_0$ &  $\bar{K}^{\mathcal{L}}_{\rm{iso}}$ & $\bar{G}^{\mathcal{L}}_{\rm{iso}}$ & $\zeta_0$  & $\bar{K}^{\mathcal{L}}_{\rm{iso}}$ & $\bar{G}^{\mathcal{L}}_{\rm{iso}}$ & $\zeta_0$\\ 	
		  	&\multicolumn{3}{c|}{Ru}&\multicolumn{3}{c|}{Ti}&\multicolumn{3}{c|}{Co}\\  
		S-128-BCC & 157.38 & 80.37 & 0.63 & 93.68 & 21.46 & 0.70  & 196.86 & 34.85 & 1.08\\
		M-250-BCC& 162.94 & 102.96  & 1.09 & 96.29 & 23.25 & 5.42 & 196.43 &  49.17 & 1.48\\
		M-128-BCC & 179.71 & 113.93 & 0.93 & 100.70 & 28.48 & 3.99  & 197.56 & 55.10 & 1.23\\
		M-125-Random & 181.50 & 96.53 & 2.34  & 98.76  & 28.14 &  2.29 & 197.05 & 48.25 & 2.93\\
		M-54-BCC& 218.25 & 114.07 & 2.30 & 100.35 & 28.22 & 5.01 &  196.63 & 60.10 & 1.14\\
		M-16-BCC & 221.48 & 134.40 & 0.91 & 103.81 & 35.73 & 1.57   & 195.61 & 65.59 & 1.61\\
		L-16-BCC& 251.27 & 155.36  & 0.65 & 107.36 & 41.33 & 0.61  & 195.06 & 75.21 & 0.86\\ \hline
			&\multicolumn{3}{c|}{Zr}&\multicolumn{3}{c|}{Mg}&\multicolumn{3}{c|}{Re}\\  
		S-128-BCC & 85.55 & 17.34  & 1.38 & 33.14 & 6.29 & 3.27  & 341.05 & 88.11 & 0.37\\
		M-250-BCC& 96.98 & 21.51 & 3.03 & 33.59  & 7.81 & 5.35 & 297.46 & 62.10 & 1.84\\
		M-128-BCC & 99.12 & 23.23 & 2.77 &33.88  & 9.22 & 2.61  & 297.04 & 66.01 & 1.08\\
		M-125-Random & 99.02  & 23.76 & 2.55 & 33.94 & 9.27 & 1.91  & 301.03 & 67.43 & 0.41\\
		M-54-BCC& 103.51  & 26.31  & 3.16 & 34.15  & 10.05 & 2.32 & 293.37 & 62.29 & 0.80\\
		M-16-BCC & 108.51 & 30.67 & 3.15 & 34.50 & 11.68 & 1.53  & 291.08 & 60.16 & 1.51\\
		L-16-BCC& 114.89 & 35.00 & 1.36 & 34.76 & 12.48 & 2.05  & 284.48 & 56.67 & 1.66\\
		\hline 
	\end{tabular}
\end{table}

\begin{table}[H] 
	\caption{Constants $K$, $G_1$, $G_2$, $G_3$ and  $L_{12}$ of monocrystal samples defining four Kelvin moduli and the stiffness distributor, identified shell elastic moduli $K_{\rm{s}}$ and $G_{\rm{s}}$ and \emph{cutoff radius} $\Delta$ of the applied atomistic potential for analysed metals. Metals are ordered with an increasing anisotropy parameter $\zeta_0$. Respective universal elastic anisotropy index $A^{\rm{U}}$ is also included.}
	\label{tab:Shell-CutOff}
	\centering
	\renewcommand{\arraystretch}{1.5}
	\footnotesize 
	\begin{tabular}{|c c c c c c c c c c c|}
		\hline
		Metal & $K$ & $G_1$ & $G_2$ &  $G_3$ &  $L_{12}$ & $\zeta_0$ & $A^{\rm{U}}$ & $K_s$ & $G_s$ & $\Delta$\\ 
	& [GPa] & [GPa] & [GPa] & [GPa] & [GPa] & [\%] & & [GPa] & [GPa] & [$\AA$]\\ \hline
	Ru & 303.9 & 211.9 & 188.3 & 199.6 & 34.65 &0.83 &0.016 & 157.4 & 80.37& 7.6 \\
	Ti & 112.2& 54.57 & 43.63 & 52.79 & 5.346 &1.91 &0.051 & 93.68 & 21.46 & 6.72\\
	Co & 194.1 & 115.5 & 82.17 & 92.54 & 10.05&2.11 &0.079 & 196.9 & 34.85 & 6.5 \\ 
	Zr & 122.4 & 64.10 & 32.28 & 46.45 & 3.757&5.00 &0.345 & 85.55 & 17.34 & 7.6 \\
	Mg & 35.48 & 23.77 & 13.59 & 13.86 & 2.362 &5.50 &0.250 & 33.15 & 6.290 & 7.15 \\
	Re& 280.1 & 104.3 & 40.15 & 52.51 & 31.30&6.29 &0.656 &341.1 & 88.11& 5.5 \\	
	\hline 
	\end{tabular}
\end{table}

\subsection{Comparison of atomistic and mean-field estimates}
\label{ssec:CompACest}

The estimates of effective bulk and shear moduli obtained by the core-shell model for nanocrystalline hcp metals are now compared with the results of atomistic simulations reported in Table \ref{tab:SamplesHCP}. All analytical estimates are calculated for perfectly random distribution of orientation, so the overall stiffness specified by Eq. (\ref{eq:core-shell}) is isotropic. 

First, let us discuss the classical bounds and mean-field estimates for coarse grained polycrystals of six hcp metals with local properties specified in Table \ref{tab:Shell-CutOff}. Their values for each metal are collected in Table \ref{tab:BoundsCM}.
As it is seen, due to small non-coaxiality angle $\Phi$, the Reuss and Voigt bounds on the bulk modulus are very close. A larger difference between those bounds exists as concerns the shear modulus. Nevertheless when one compares the value of the self-consistent and CS/MT estimates they are again close to each other. This observation leads to the conclusion that the estimates delivered by two variants of a core-shell model for nanocrystalline medium will not be far from each other as well. Therefore, since the CS/MT estimate is specified by an explicit and closed form equation, contrary to the implicit CS/SC estimate, the analysis of the model validity is focused on this variant. For this range of grain sizes atomistic simulations are not applicable due to hardware limitations related to excessively large number of atoms required to represent polycrystal. Instead, in Table \ref{tab:BoundsCM} for a purpose of comparison, results of computational FE homogenization \cite{Bohlke10} are included. Effective properties have been obtained using RVE geometries and periodic boundary conditions described in \cite{Frydrych18}. For each hcp metal 5 realizations of RVE composed of 125 grains with randomly selected orientations and $6^3$ elements per grain were analyzed to find the effective elasticity tensor $\bar{\mathbb{C}}^{FE}$. Isotropized bulk and shear moduli of such tensor are reported in Table \ref{tab:BoundsCM}. It is seen that the obtained values are close to 
the SC estimate. This result is in agreement with other literature studies, e.g. \cite{Bohlke10,Kowalski16}.


\begin{table}[!htp]
	\caption{The overall bulk and shear modulus $\bar{K}^{\infty}_{\rm{iso}}$ and $\bar{G}^{\infty}_{\rm{iso}}$\,[GPa] of coarse-grained polycrystal, obtained by the Voigt (V), Reuss (R), self-consistent (SC) estimate, the limit value obtained by MT core-shell (CS/MT) model (Eq. (\ref{Eq:KsCSMTlim}) and (\ref{Eq:GsCSMTlim})) and computational FE homogenization for polycrystals with random orientation distributions and six metals of hexagonal symmetry. Local properties of single crystal are collected in Table \ref{tab:Shell-CutOff}.
	}
     \label{tab:BoundsCM}\vspace{.05in}
     \centering
     \tiny
	\begin{tabular}{|c|ccccc|ccccc|}
		\hline
		Metal & R & SC &   
		V & CS/MT & {FE} & R & SC & V & CS/MT & {FE} \\ 
		&\multicolumn{5}{c|}{ $\bar{K}^{\infty}_{\rm{iso}}$[GPa]}&\multicolumn{5}{c|}{ $\bar{G}^{\infty}_{\rm{iso}}$[GPa]}\\\hline	
		Ru & 302.98 & 303.44 & 303.92 & 303.42 & 303.44& 197.04 & 197.28 & 197.54 & 197.18 & 197.29\\ 
		Ti & 112.095 & 112.14 & 112.18 & 112.12 & 112.14 & 48.98 & 49.24 & 49.47 & 49.16 & 49.25\\
		Co & 193.94 & 194.01 & 194.09 & 193.98 & 194.01 & 91.54 & 92.25 & 92.98 & 91.97 & 92.28\\
		Zr & 122.35  & 122.365 & 122.39 & 122.36 & 122.37 & 41.455 & 42.92 & 44.31 & 42.36 & 42.98\\
		Mg & 35.44 & 35.45 & 35.48 & 35.45 & 35.46  & 14.99 & 15.32 & 15.74 & 15.19 & 15.33\\
		Re & 278.52  & 279.14 & 280.085 & 279.31 & 279.16 & 51.255 & 54.24& 57.92& 54.96 & 54.41\\
		\hline
	\end{tabular}\\[.051in]
\end{table}

\begin{figure}
	\centering
	\begin{tabular}{cc}
		(a) & (b)\\
		\includegraphics[angle=0,width=0.49\textwidth]{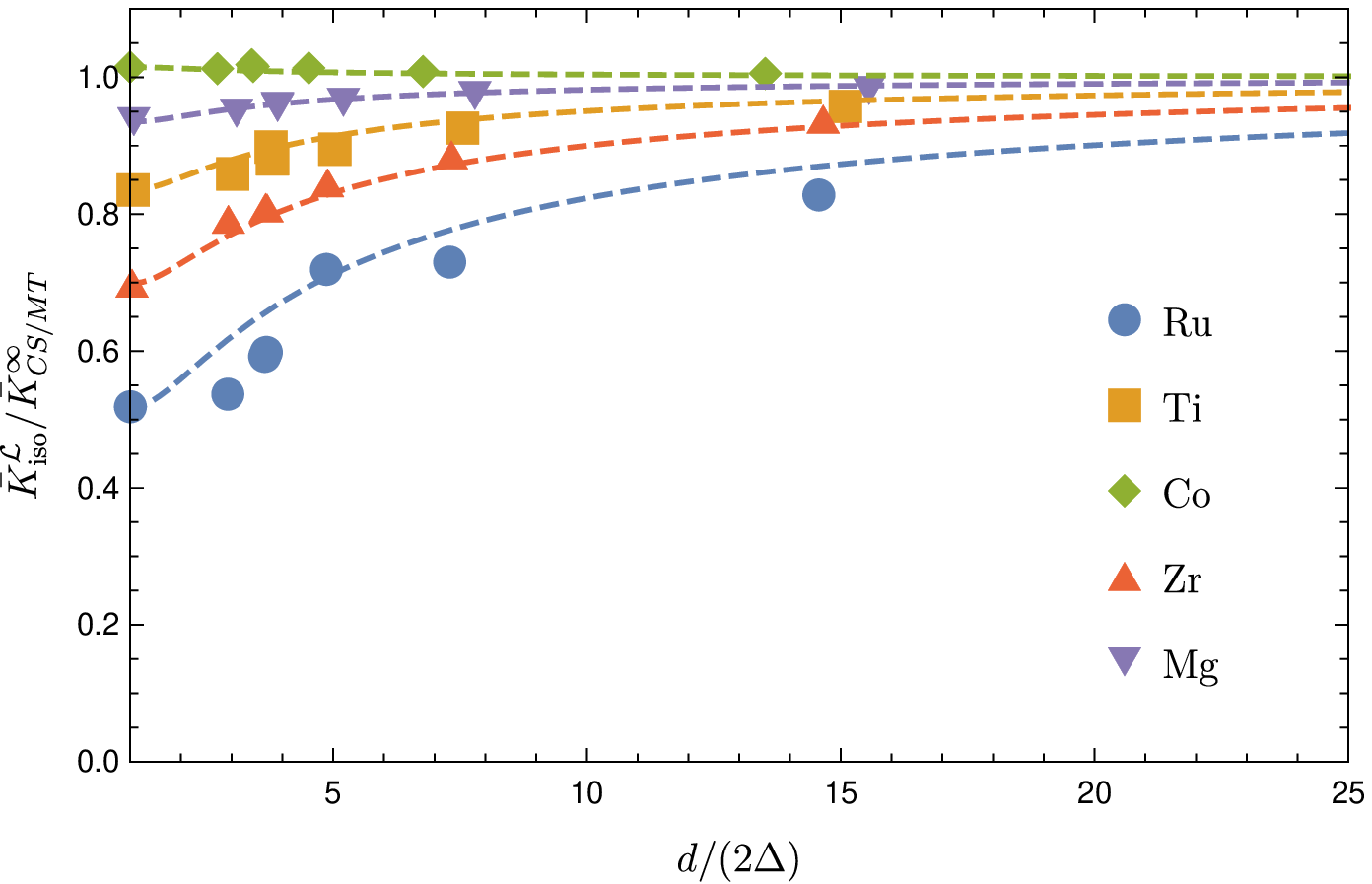}&\includegraphics[angle=0,width=0.49\textwidth]{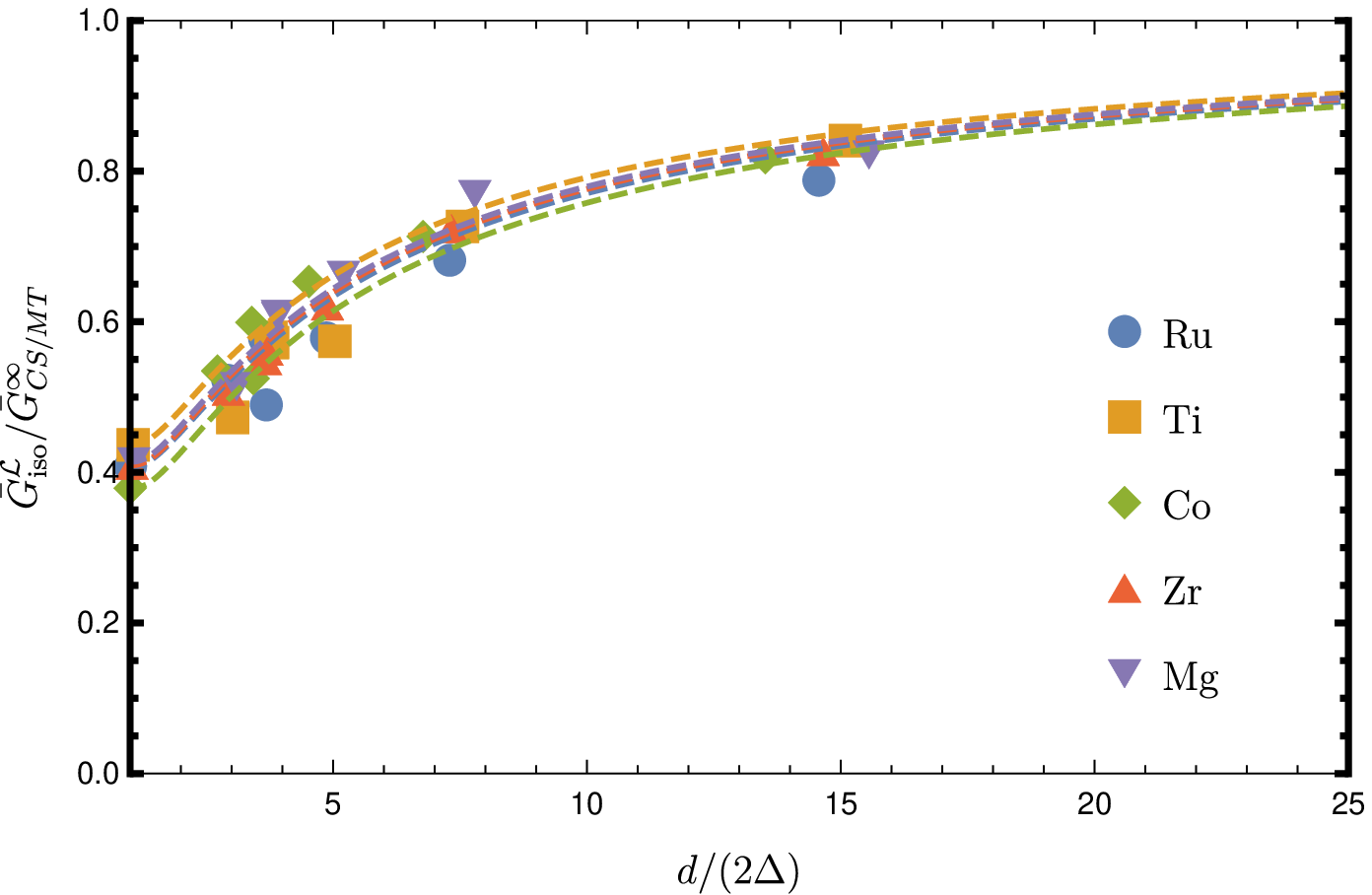}\\
		(c) & (d)\\
		\includegraphics[angle=0,width=0.49\textwidth]{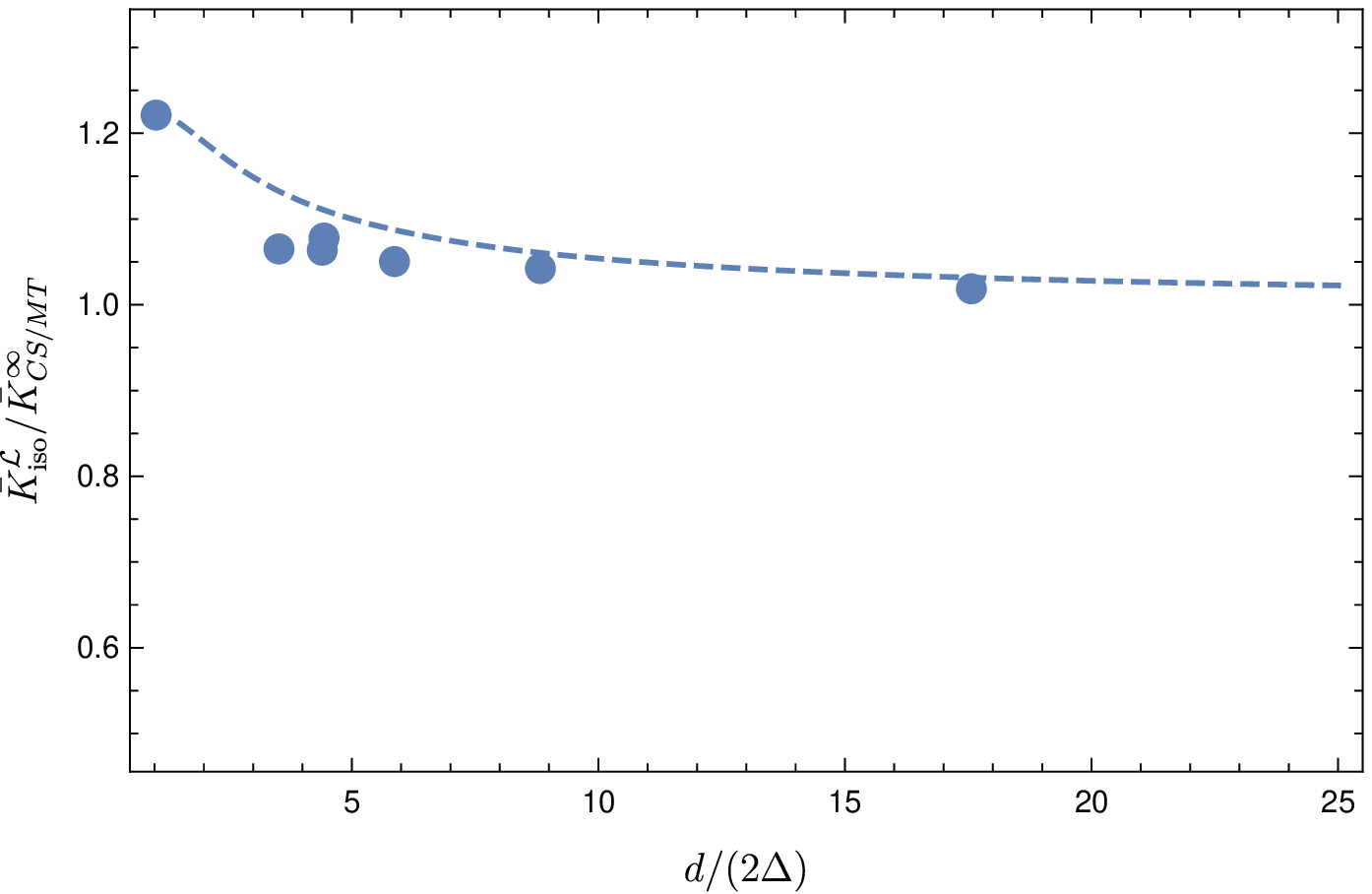}&\includegraphics[angle=0,width=0.49\textwidth]{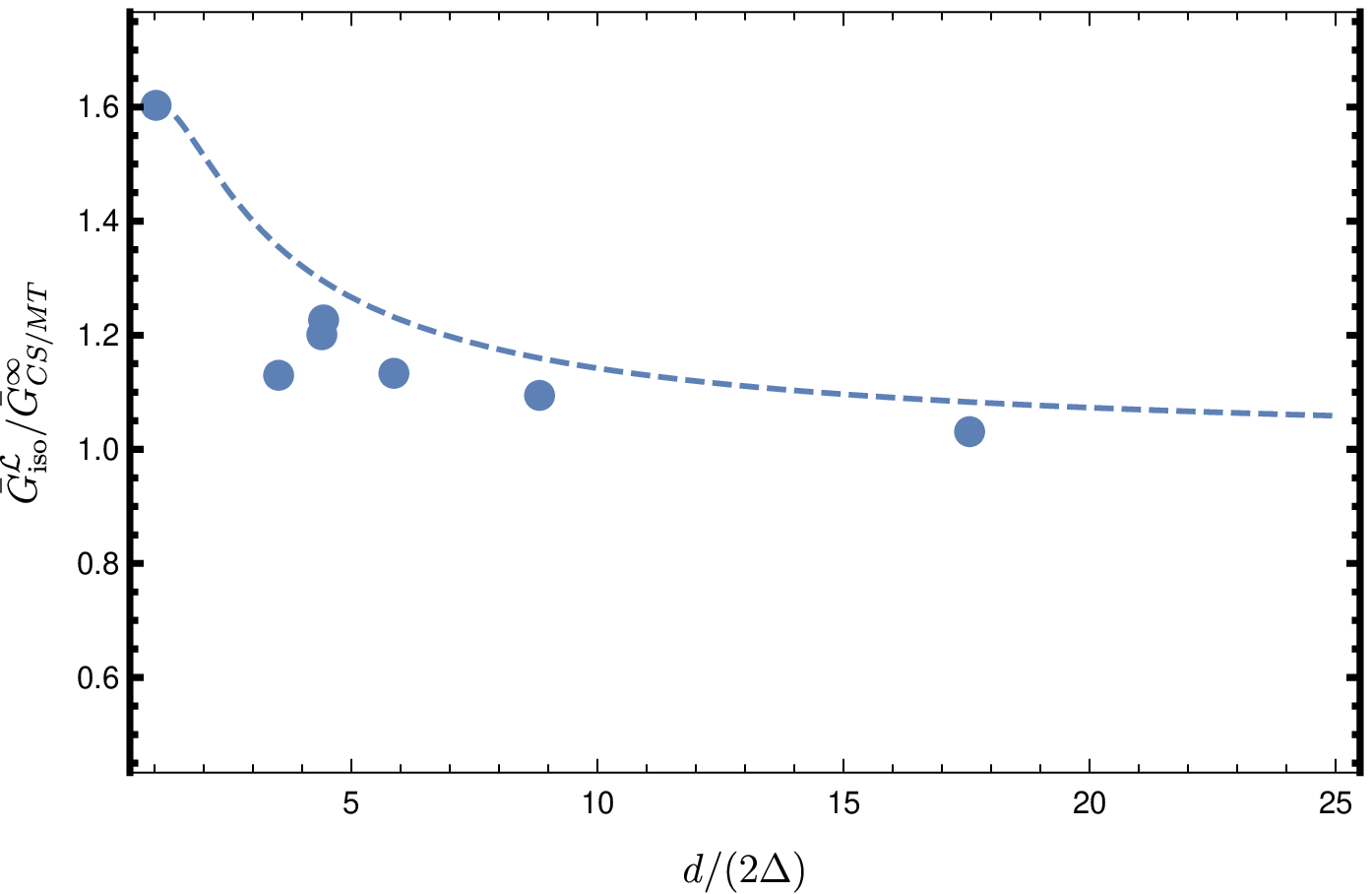}
	\end{tabular}
	\caption{The isotropic bulk $\bar{K}^{\mathcal{L}}_{\rm{iso}}$ (a) and shear $\bar{G}^{\mathcal{L}}_{\rm{iso}}$ (b) moduli as functions of the average grain diameter $d$ for 5 hcp metals: Ru, Ti, Co, Zr and Mg - the results of atomistic simulations reported in Tables \ref{tab:Cij-cRu}, \ref{tab:Cij-cTi}, \ref{tab:Cij-cCo}, \ref{tab:Cij-cZr} and \ref{tab:Cij-cMg} as well as the Mori-Tanaka core-shell (CS/MT) model predictions (dashed line); (c) and (d) contain analogical plots for Re (Table \ref{tab:Cij-cRe}). For the purpose of comparison the grain diameter is scaled by the double \emph{cutoff radius} of atomistic potential $2\Delta$ for the given metal (the last column in Table \ref{tab:Shell-CutOff}), while moduli are scaled by the respective estimates of CS/MT scheme for the coarse-grained random polycrystal (Table \ref{tab:BoundsCM}).}\label{Fig:5mGK} 
\end{figure}

Figure \ref{Fig:5mGK} compares the CS/MT model predictions with the corresponding results of atomistic simulations for nano-grained polycrystals. Since there is a huge difference in the elastic stiffness between the analyzed metals (e.g. the shear modulus of the coarse-grained Mg polycrystal is almost 13 times smaller than for Ru), in order to study the general trends, the dimensionless quantities are used. The moduli for nanocrystalline materials are scaled by the respective values for a coarse-grained polycrystal: $\bar{K}_{CS/MT}^{\infty}$ and $\bar{G}_{CS/MT}^{\infty}$ (Table \ref{tab:BoundsCM}), while the grain diameter by a double \emph{cutoff radius} of corresponding atomistic potential. As it is seen in Fig. \ref{Fig:5mGK}b the shear modulus of five out of six hcp metals (Ru, Ti, Co, Zr, Mg) follows the common qualitative and quantitative trend -- with a decreasing grain diameter the value drops from $\bar{G}_{CS/MT}^{\infty}$ to the value of approximately $0.4\bar{G}_{CS/MT}^{\infty}$ when the grain boundary zone encompasses the whole volume. For a grain diameter of $30\times\Delta$ the value of  $0.8\bar{G}_{CS/MT}^{\infty}$ is attained. As concerns bulk modulus the qualitative trend is similar, however, quantitatively the relative value attained when $d/(2\Delta)\rightarrow 1$ varies between metals from almost 1 for Co (i.e. very little variation of the bulk modulus with a grain size) to 0.5 for Ru (relatively strong variation). The core-shell model predicts this behaviour quite accurately. For Ren an opposite qualitative trend is observed, i.e. an increase of two moduli with a decrease of grain size. For this metal the bulk and shear modulus of a boundary zone established in atomistic simulations are larger than effective properties of the random coarse grained polycrystal. Nevertheless, also in this case, the CS/MT model estimates are in a good agreement with atomistic calculations.    

\begin{figure}
	\centering
	\begin{tabular}{ccc}
		&Young's modulus &  Poisson's ratio\\
		a) \textbf{Ru}&
		\includegraphics[angle=0,width=0.39\textwidth]{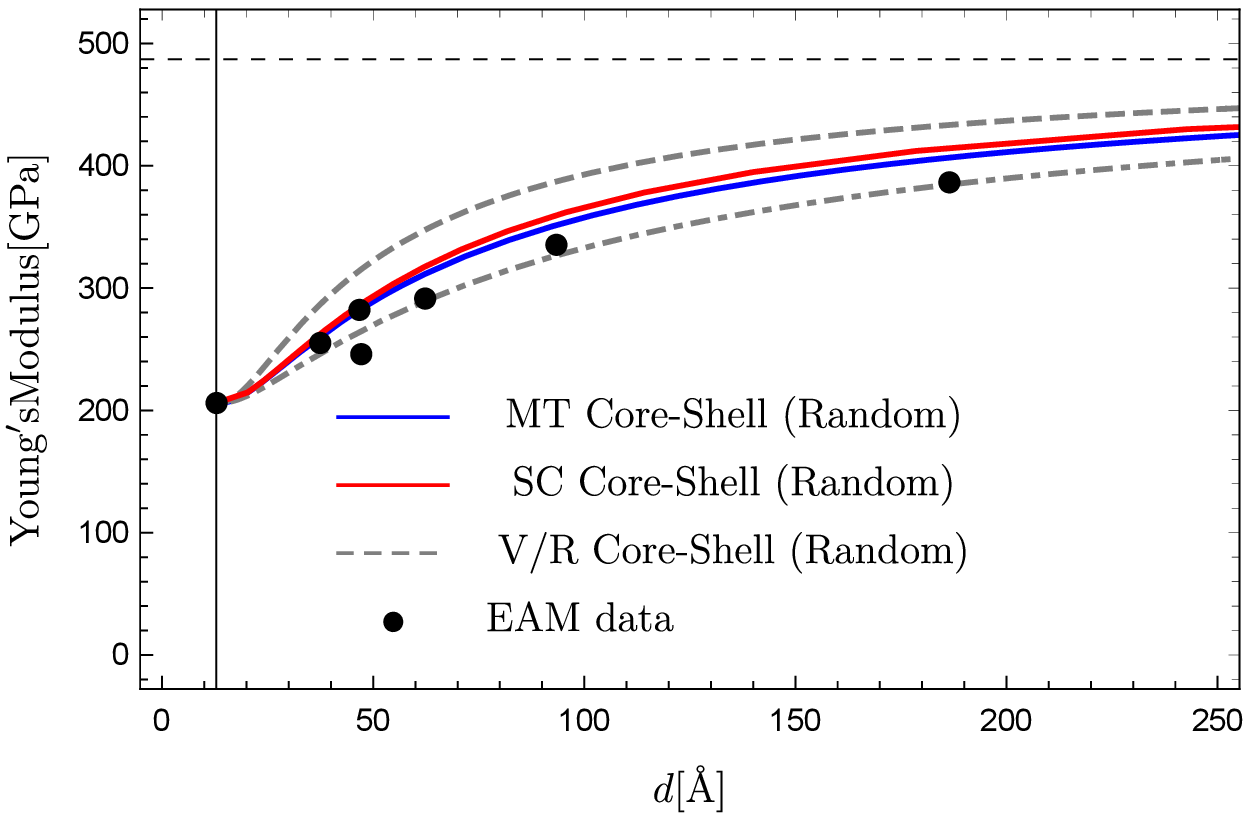}
		&
    	\includegraphics[angle=0,width=0.39\textwidth]{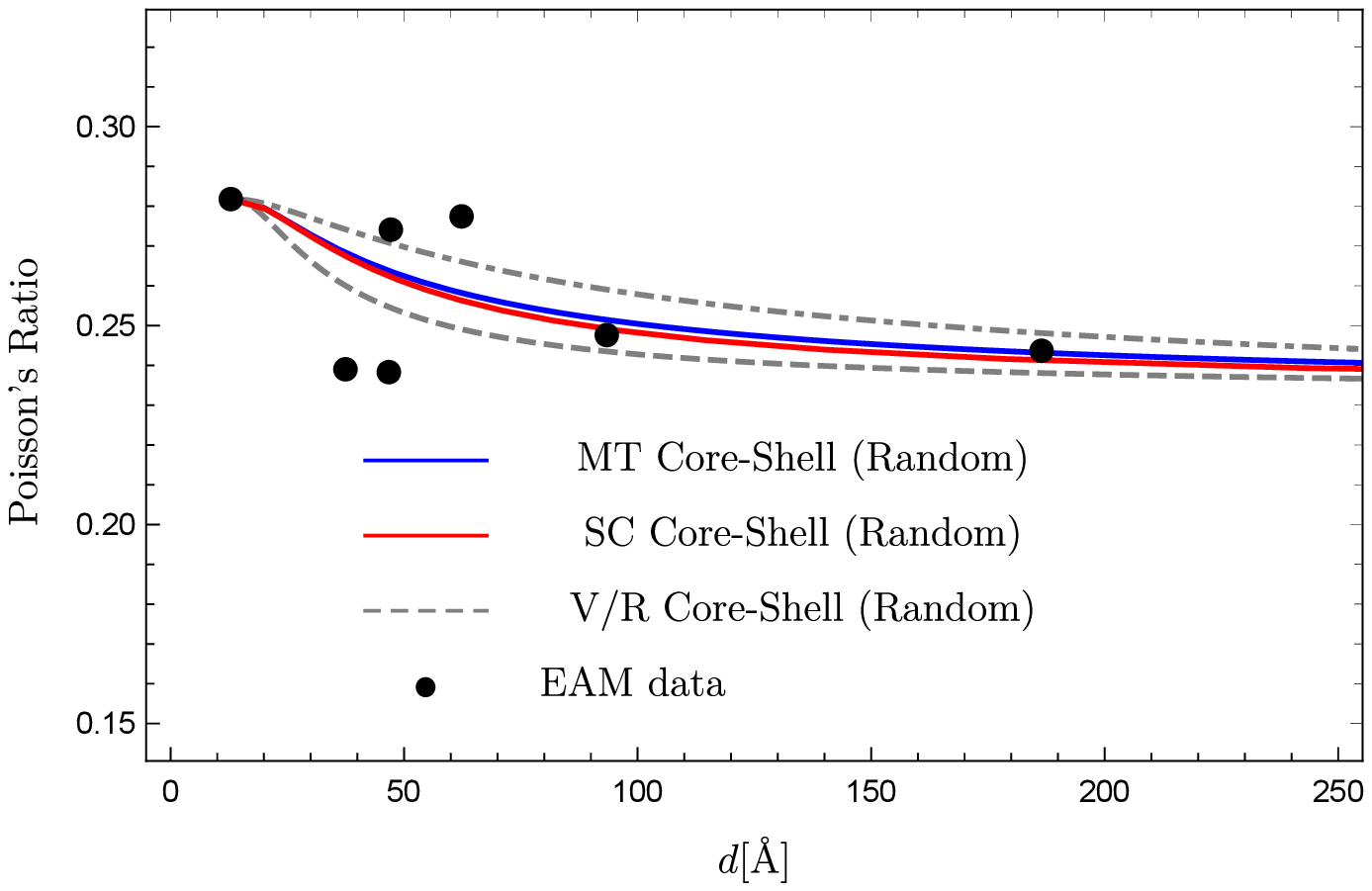}
		\\
		b) \textbf{Ti}&	\includegraphics[angle=0,width=0.39\textwidth]{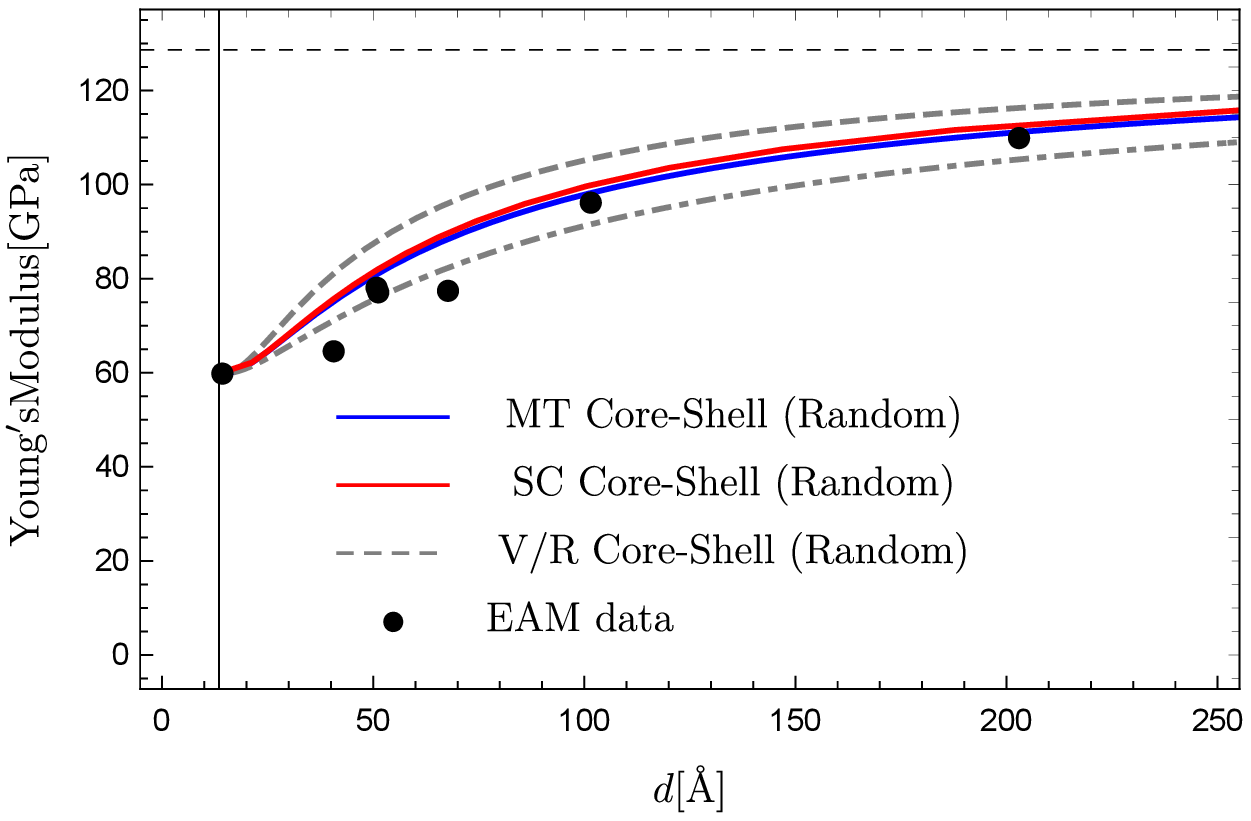}
		&
		\includegraphics[angle=0,width=0.39\textwidth]{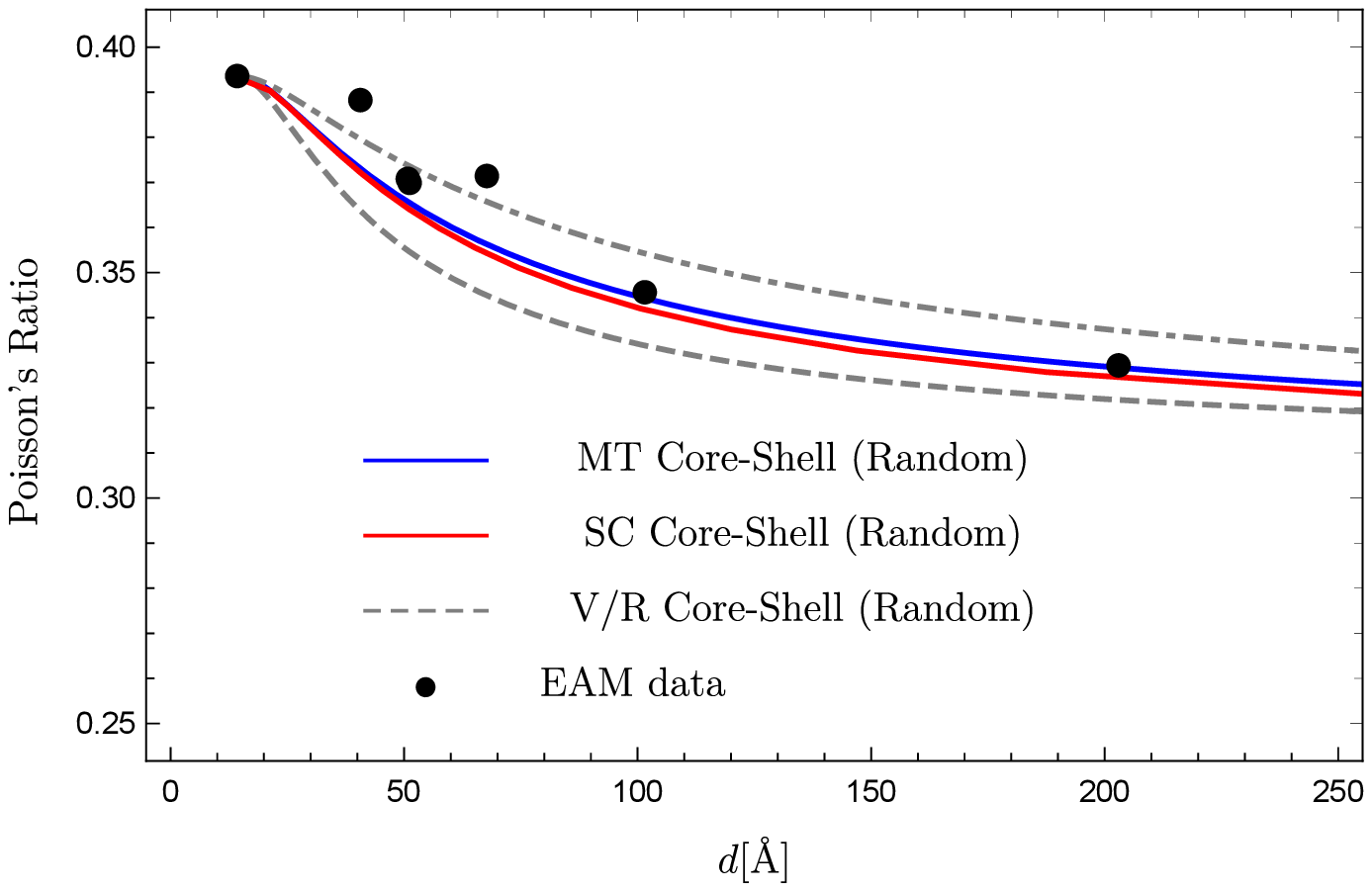}
		\\
		c) \textbf{Co}&	\includegraphics[angle=0,width=0.39\textwidth]{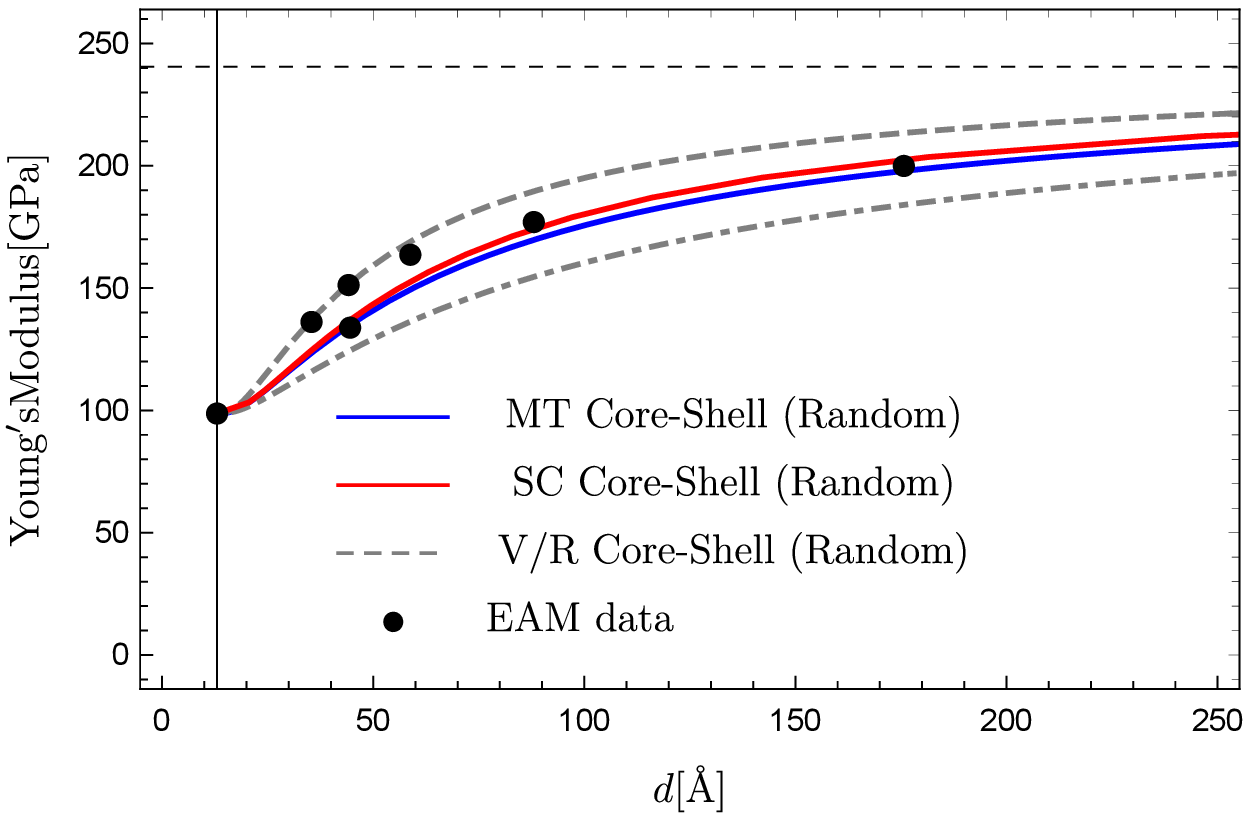}
		&
		\includegraphics[angle=0,width=0.39\textwidth]{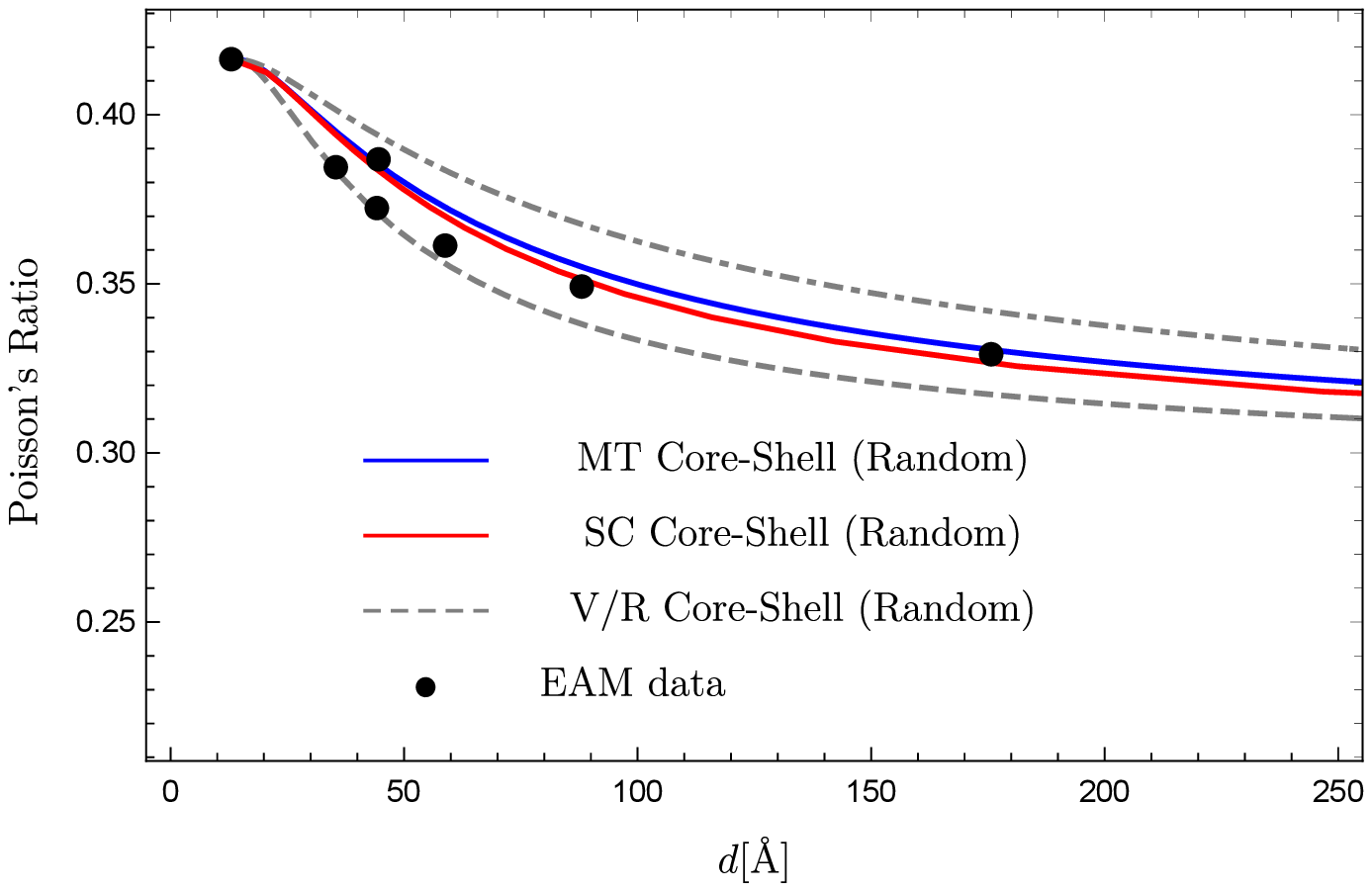}
		\\
	\end{tabular}
	\caption{The isotropic Young modulus $\bar{E}^{\mathcal{L}}_{\rm{iso}}$ and Poisson's ratio  $\bar{\nu}^{\mathcal{L}}_{\rm{iso}}$ as a function of the average grain diameter $d$ by the two variants of the core-shell model (CS/MT, CS/SC) and the two-phase iso-strain Voigt (V) and iso-stress Reuss (R) schemes - comparison with results of atomistic simulations, calculated using Eq. \ref{Eq:E-nu}:  (a) Ru, (b) Ti (c) Co (d) Zr (e) Mg (f) Re. The horizontal dashed black lines in left figures indicate the limit value of CS/MT estimate of Young's modulus for a coarse-grained polycrystal.} 
	\label{fig:ZetaE1}
\end{figure}

The quantitative comparison of the proposed mean-field model predictions and the results of atomistic simulations concerning the overall Young modulus and Poisson's ratio for six hcp metals is demonstrated in Fig. \ref{fig:ZetaE1}. Presented values are calculated using the well-known relations: 
\begin{equation}\label{Eq:E-nu}
\bar{E}^{\mathcal{L}}_{\rm{iso}}=\frac{9\bar{K}^{\mathcal{L}}_{\rm{iso}}\bar{G}^{\mathcal{L}}_{\rm{iso}}}{3\bar{K}^{\mathcal{L}}_{\rm{iso}}+\bar{G}^{\mathcal{L}}_{\rm{iso}}}\,,\quad\bar{\nu}^{\mathcal{L}}_{\rm{iso}}=\frac{3\bar{K}^{\mathcal{L}}_{\rm{iso}}-2\bar{G}^{\mathcal{L}}_{\rm{iso}}}{6\bar{K}^{\mathcal{L}}_{\rm{iso}}+2\bar{G}^{\mathcal{L}}_{\rm{iso}}}\,.
\end{equation} 
The Young modulus follows qualitatively the trend observed for the shear modulus. As concerns Poisson's ratio for Ru, Ti, Co, Zr and Mg, it decreases with a grain size, while an opposite relation is found for Re. Additionally, presented results confirm the observation that for analyzed hcp metals the CS/MT and CS/SC estimates are close to each other. In spite of these two mean-field models figures contain also predictions obtained using the mixture rule-based iso-strain (Voigt) scheme and its counterpart -- an iso-stress Reuss scheme. Those two are upper and lower bounds for stiffness moduli (but not Poisson's ratio) of a two-phase random polycrystalline medium. Comparing the predictions of all presented averaging models with the atomistic simulations it is seen that on overall the CS/MT scheme can be recommended as delivering reasonable predictions for all hcp metals. Moreover, consistency of model estimates with the results of atomistic simulations proves validity of the assumed procedure for the assessment of size and average properties for the grain boundary zone.     

\setcounter{figure}{5}

\begin{figure}
	\centering
	\begin{tabular}{ccc}
		&Young's modulus &  Poisson's ratio\\
		d) \textbf{Zr}& \includegraphics[angle=0,width=0.39\textwidth]{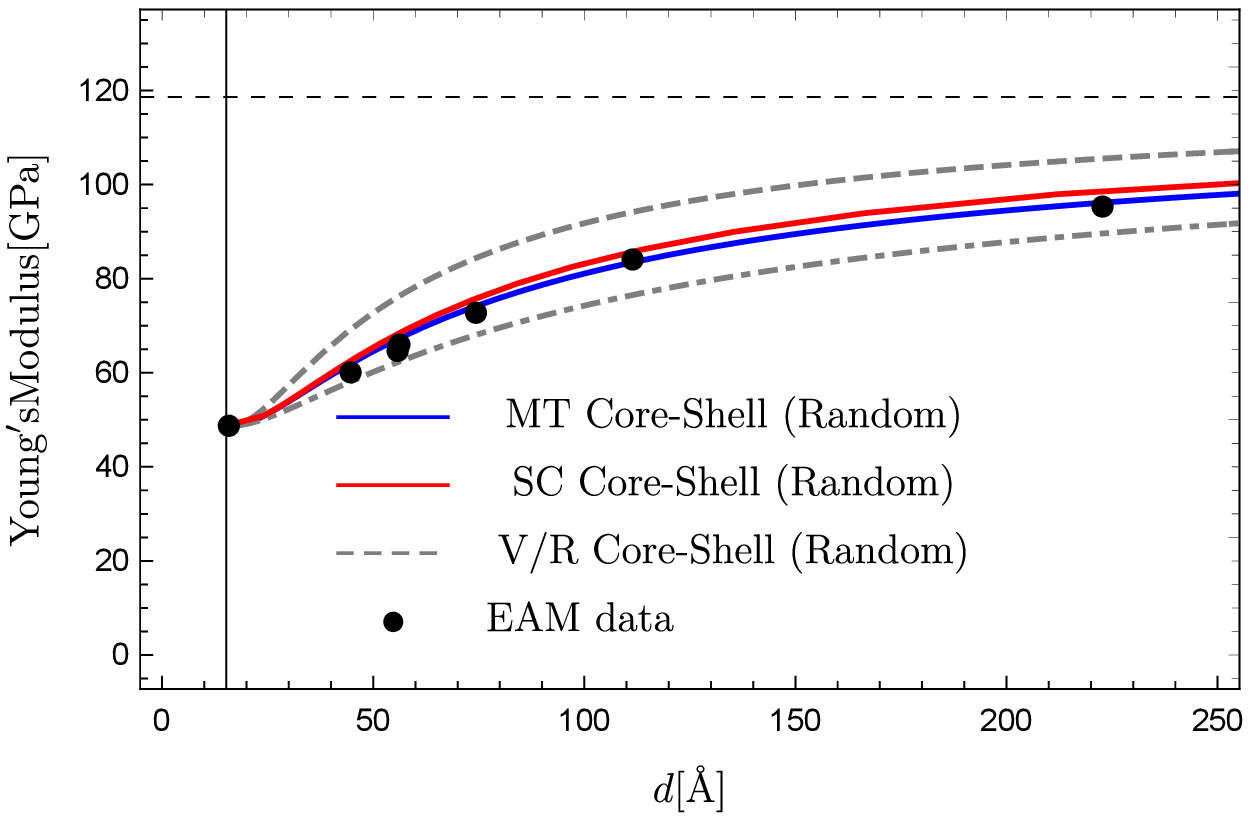}
		&
		\includegraphics[angle=0,width=0.4\textwidth]{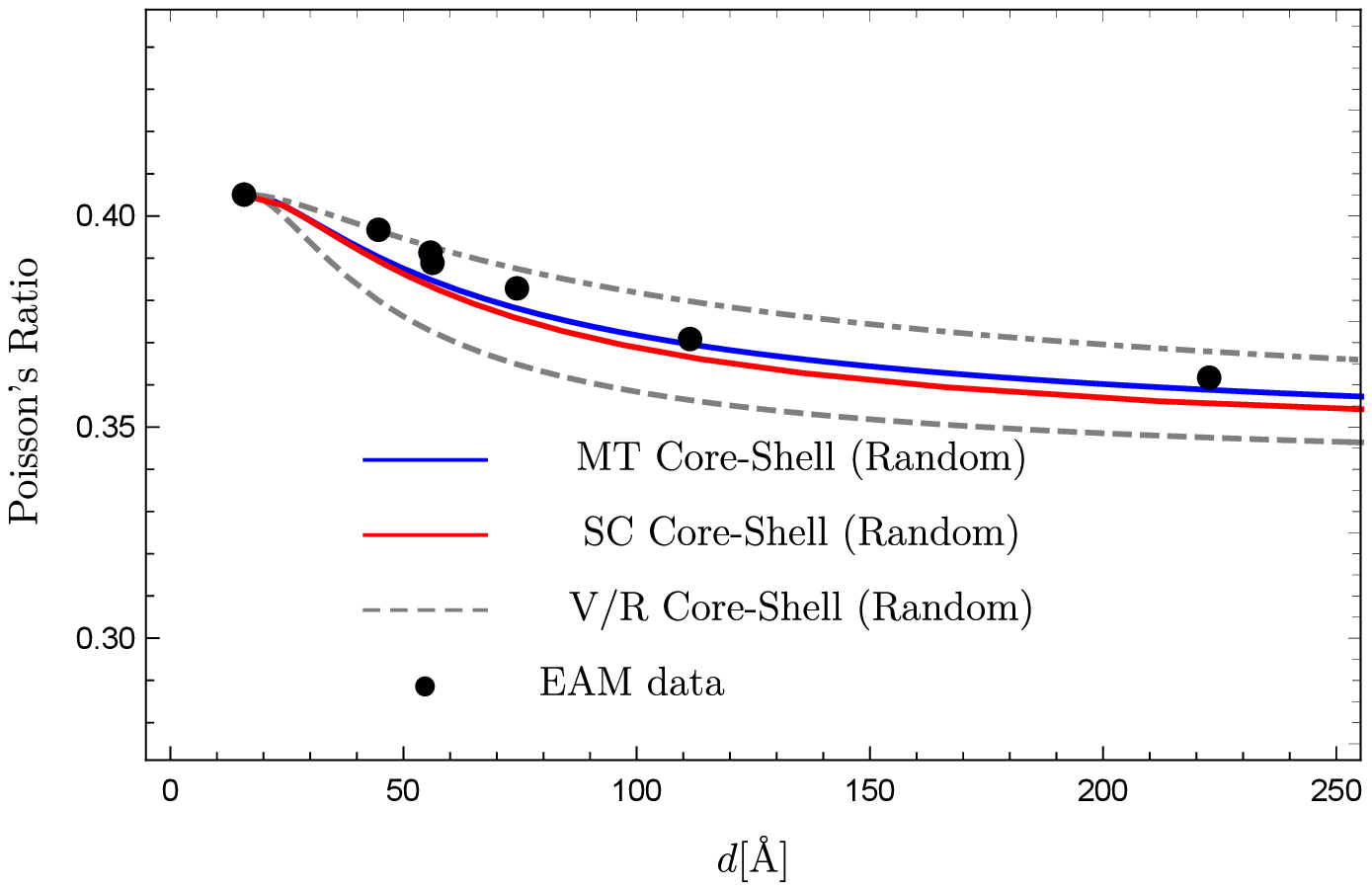}
		\\
		e) \textbf{Mg}& \includegraphics[angle=0,width=0.39\textwidth]{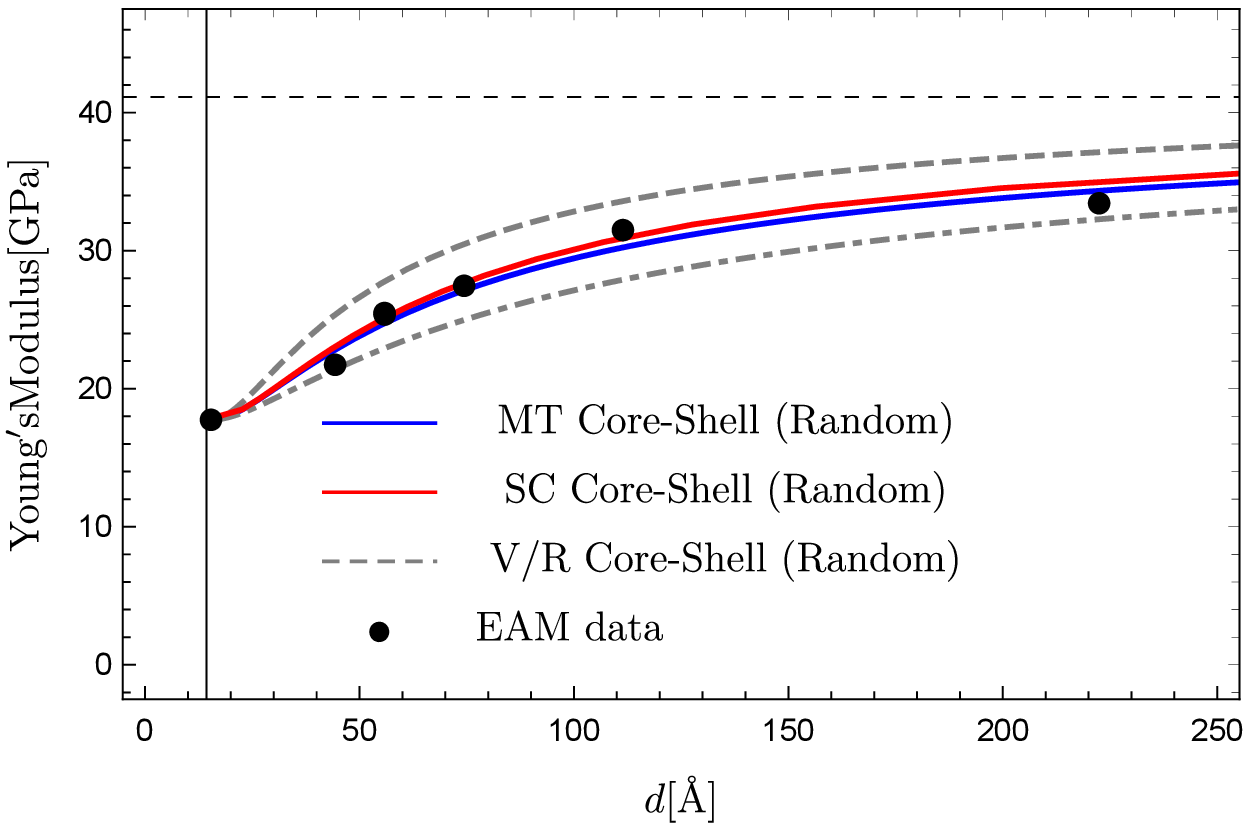}
		&
		\includegraphics[angle=0,width=0.39\textwidth]{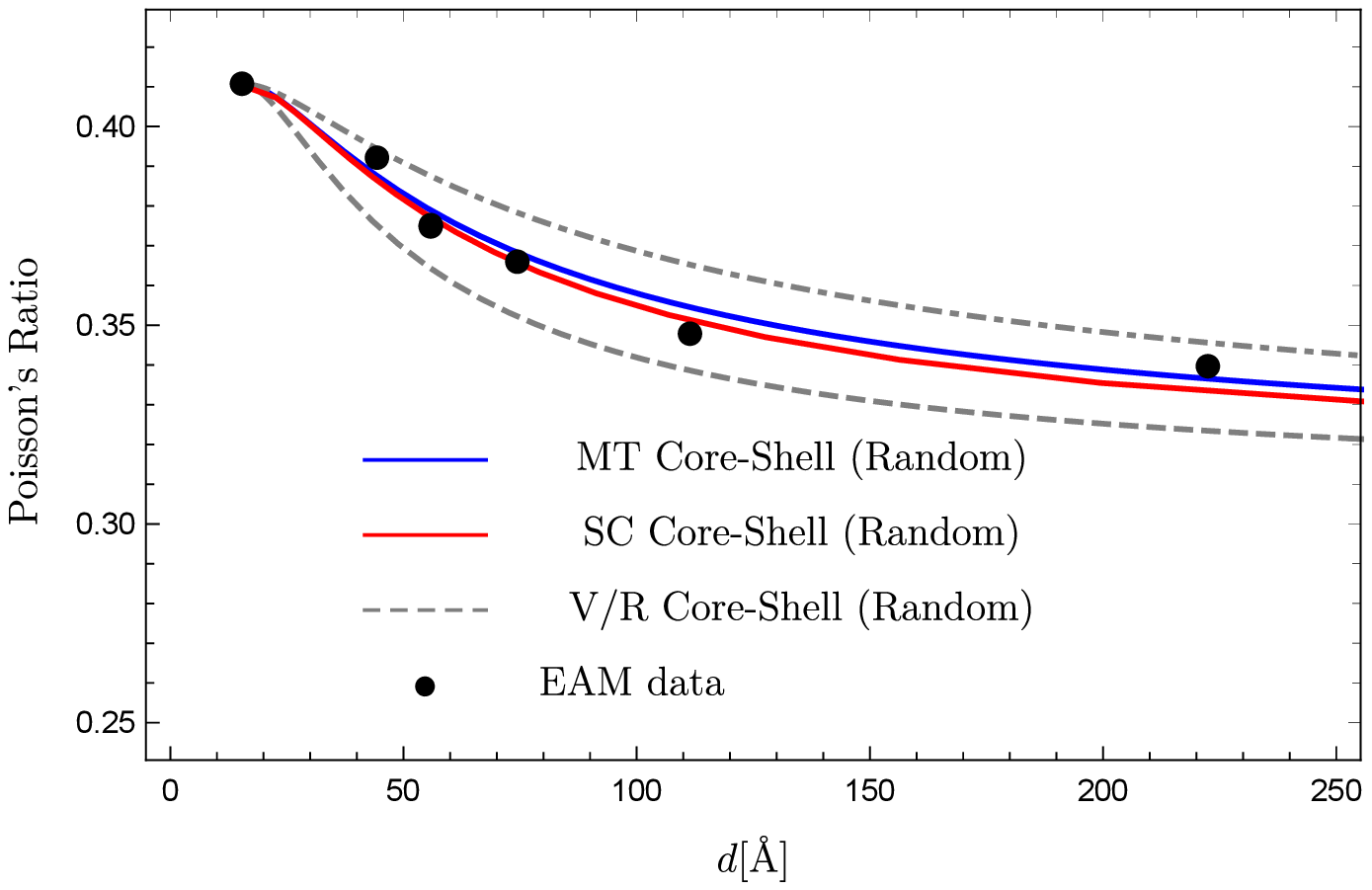}
		\\
		f) \textbf{Re}& \includegraphics[angle=0,width=0.39\textwidth]{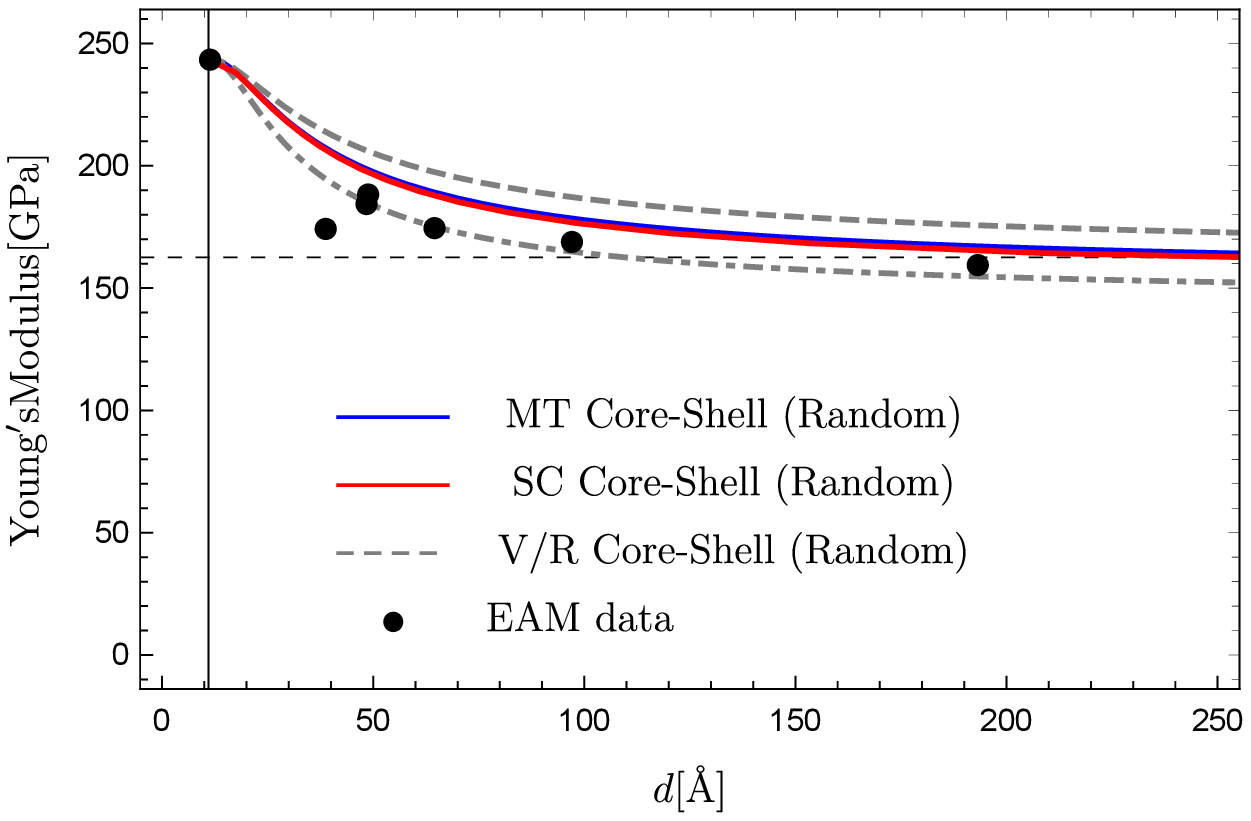}
		&
		\includegraphics[angle=0,width=0.39\textwidth]{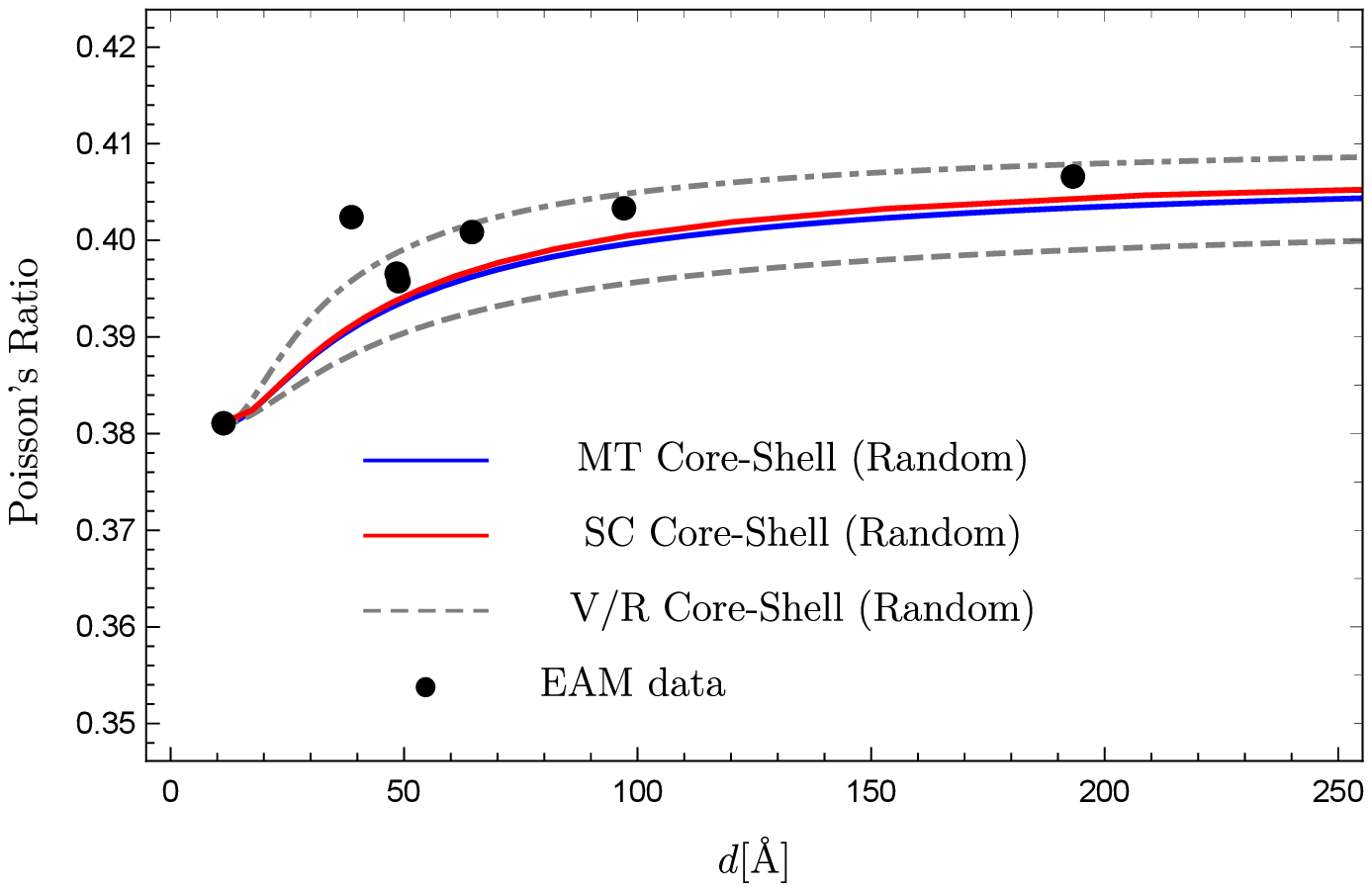}
		\\
	\end{tabular}
	\caption{continue...} 
	
\end{figure}

In \cite{Kowalczyk19} the correlation between the Zener parameter and the character of the relation between effective moduli and the grain size has been found for cubic nanocrystalline metals. However, in the case of analyzed six hcp metals it is difficult to indicate similar correlation between the anisotropy degree and the grain size effect. Although Re shows the highest anisotropy degree quantified by $\xi_0$ and $A^U$ (see Table \ref{tab:Shell-CutOff}), it is not that much different from other five metals under study. From this point of view, it would be interesting to verify if the identified behaviour is an artefact consequence of the inherent features of the applied atomistic potential or is also observed in reality. Unfortunately, authors were not able to find any experimental data in the literature to confirm either of hypotheses.

\section{Summary and conclusions}
\label{sec:Con}

{Different variants of a mean-field core-shell model \cite{Kowalczyk18,Kowalczyk19} for estimation of elastic properties of bulk nanocrystalline metals have been validated for hcp crystal lattice symmetry. Because there is not enough experimental data, validation has been conducted by comparing the estimates with the results of atomistic simulations. Six metals of hexagonal (hcp) lattice geometry were selected for which the verified EAM potentials are available in the literature. All of them are characterized by relatively low non-coaxiality angle $\Phi$ and the same relation between Zener-like anisotropy factors (\ref{Eq:zeta1}),
see also the collective figures \ref{fig:Zener}.}   

Following previous research \cite{Kowalczyk18,Kowalczyk19}, for each hcp metal atomistic simulations have been conducted on seven generated samples of polycrystalline materials with randomly selected orientations. Samples vary as concerns the average grain size, so that the averaged grain diameter takes values between ca. 1 nm to 20 nm. In the simulations all 21 components of the anisotropic elastic stiffness tensor are identified. The smallest sample served to identify the average  properties of the grain boundary zone.  For further analysis of the grain-size effect on the elastic moduli the closest isotropic approximation is found using the Log-Euclidean norm \cite{Moakher06}. 

It has been observed that for five out of six studied metals (Ru,Ti,Co,Zr,Mg) the elastic bulk and shear moduli increase with a grain size. The reverse trend is observed for rhenium (Re). This metal exhibits the strongest anisotropy among considered metals, although the correlation between the anisotropy degree and the character of grain size dependence is not clear. It would be interesting to confirm experimentally this qualitative difference in the grain size effect for this metal, since the present observations strongly relays on validity of the applied atomistic potential.     

Among the considered variants of core-shell model the estimates of elastic moduli obtained by the Mori-Tanaka scheme are on overall in the most satisfactory qualitative and quantitative agreement with the results of atomistic simulations for all considered hexagonal metals, independently of the character of the grain size effect. The study demonstrated also the validity of the assumptions concerning  the shell thickness and properties.

The applied variants of mean-field core-shell model can be extended to estimate a non-linear response of a nanocrystalline material and specifically the yield strength \cite{Jiang04,Capolungo07}. Atomistic simulations may serve to validate such an extension.

\section*{ACKNOWLEDGMENTS}
The research was partially supported by the project No. 2016/23/B/ST8/03418 of the National Science Centre, Poland.
Additional assistance was granted through the computing cluster GRAFEN at Biocentrum Ochota, the
Interdisciplinary Centre for Mathematical and Computational Modelling of Warsaw University (ICM UW) and Pozna\'n Supercomputing and Networking
Center (PSNC). 
 
\appendix

\section{Spectral decomposition of elasticity tensor for hcp crystal\label{spek-ti} and standard estimates of effective stiffness for random polycrystals of hexagonal symmetry\label{Ap:1}}

Spectral decomposition of $\mathbb{C}$ for a crystal of hcp symmetry is given by Eq. (\ref{Eq:SpekTrans}), where the projectors $\mathbb{P}_3(\phi_c)$ and $\mathbb{P}_4(\phi_c)$ for two 2D eigen-subspaces are given by Eq. (\ref{Eq:P3}) and (\ref{Eq:P4}). These two projectors are common for all crystal of this symmetry. Two remaining projectors depend also on the distributor
$\xi=1/3\tan\Phi$, which is material-specific, namely
\begin{eqnarray}\label{Pti1}
\mathbb{P}_1(\xi,\phi_c)\!\!&\!\!=\!\!&\!\!(\cos\Phi)^2\mathbb{I}^{\rm{P}}+
\frac{1}{2\sqrt{3}}\sin 2\Phi(\mathbf{I}\otimes\mathbf{D}_{\mathbf{n}}+\mathbf{D}_{\mathbf{n}}\otimes\mathbf{I})+(\sin\Phi)^2\mathbf{D}_{\mathbf{n}}\otimes\mathbf{D}_{\mathbf{n}}\\ \label{Pti2} 
\mathbb{P}_2(\xi,\phi_c)\!\!&\!\!=\!\!&\!\!(\sin\Phi)^2\mathbb{I}^{\rm{P}}-
\frac{1}{2\sqrt{3}}\sin 2\Phi(\mathbf{I}\otimes\mathbf{D}_{\mathbf{n}}+\mathbf{D}_{\mathbf{n}}\otimes\mathbf{I})+(\cos\Phi)^2\mathbf{D}_{\mathbf{n}}\otimes\mathbf{D}_{\mathbf{n}}\,,
\end{eqnarray}
Angle $\Phi=\Phi(\xi)$ is calculated using the components of $2\times 2$ matrix (\ref{Eq:L2x2}) as follows 
\begin{equation}\label{Eq:angleFI}
\Phi=\frac{1}{2}\arctan\left(\frac{2 L_{12}}{3K-2G_1}\right)\in \left<-\pi/4,\pi/4\right>
\end{equation} 
This angle is a measure of non-coaxiality between the given anisotropic stiffness of hcp crystal and any isotropic tensor. If $\Phi=0$ then they are coaxial and $\mathbb{C}(\phi^c)\bar{\mathbb{C}}_{\rm{iso}}-\bar{\mathbb{C}}_{\rm{iso}}\mathbb{C}(\phi^c)=\mathbb{O}$ for any $\bar{\mathbb{C}}$ specified by Eq.~(\ref{Eq:Ciso}). It is worth noting that the stiffness distributor $\xi$ can be expressed by the invariants of orthogonal projector $\mathbb{P}_1$ \cite{Kowalczyk04a}.

The formulas for standard estimates of effective elastic stiffness of one-phase polycrystals of any anisotropy has been provided in Appendix A of \cite{Kowalczyk18}. Their specification for materials of random texture composed of grains of hexagonal symmetry are collected in Table \ref{tab:LinearEstimates} (for details see \cite{Berryman05} and \cite{Kowalczyk12}). It is seen that if $L_{12}=0$, which is equivalent to $\Phi=0$, then all estimates of the overall bulk modulus coincide and are equal to $K$. Such crystals belong to the class of volumetrically isotropic materials \cite{Kowalczyk09}. It is worth to note that for hexagonal crystals and random orientation distribution the specification of two equations enabling to find the self-consistent estimate, proposed by \cite{Kroner58},  was first given by \cite{Kneer63}. It can be verified that the respective two equations in Table \ref{tab:LinearEstimates} are equivalent to Kneer's formulas.

\begin{table}[H]
	\caption{Classical mean-field estimates of the overall bulk and shear moduli ($\bar{K}$,  $\bar{G}$) for a one phase random polycrystal of hexagonal symmetry. V -- Voigt, R -- Reuss, H-S -- Hashin-Shtrikman (U -- upper, L -- lower), SC -- self-consistent (equiaxial, spherical shape of grains is assumed). $K_0^{U/L}$ and $G_0^{U/L}$ for the Hashin-Shtrikman bounds are established from the optimality conditions (for details see \cite{Berryman05} or \cite{Kowalczyk12})
	}
	\label{tab:LinearEstimates}\vspace{.05in}
	\centering
	\small{
		\begin{tabular}{ccc}
			\hline
			Estimate    &  $\bar{K}$ & $\bar{G}$ \\
			
			\hline
			V    & $K$ &  $\frac{1}{5}(G_1+2G_2+2G_3)$ \\
			\hline
			&&\\
			R & $K-\frac{L_{12}^2}{6G_1}$   &    $\left(\frac{1}{5}\left(\frac{1}{G_1-L_{12}^2/(6K)}+\frac{2}{G_2}+\frac{2}{G_3}\right)\right)^{-1}$   \\
			&&\\
			\hline
			&&\\
			H-S (U/L) & {$K-\frac{L_{12}^2}{6G_1+9K^{\small{U/L}}_{*o}}$} & 
			$\left(\frac{1}{5}\left(\frac{1}{G
				_1+G_{*o}-\frac{L_{12}^2}{6(K_1+K_{*o})}}+\frac{2}{G
				_2+G_{*o}}+\frac{2}{G
				_3+G_{*o}}\right)\right)^{-1}\!\!-\!G_{*o}$  \\ &&\\  & {\small{$K_{*o}^{U/L}=4G_{o}^{U/L}$}} & {\small{$G_{*o}^{U/L}=G_o^{U/L}\frac{9K_0^{U/L}+8G_o^{U/L}}{6(K_o^{U/L}+2G_o^{U/L})}$}}       \\ 
			&&\\
			\hline
			&&\\
			&\multicolumn{2}{c}{positive solutions of the set of two equations:}\\
			
			SC & \multicolumn{2}{c}{$K-\bar{K}-\frac{L_{12}^2}{6G_1+9\bar{K}_{*}}=0$}     \\
			&\multicolumn{2}{c}{$\left(\frac{1}{5}\left(\frac{1}{G
					_1+\bar{G}_{*}-\frac{L_{12}^2}{6(K_1+\bar{K}_{*})}}+\frac{2}{G
					_2+\bar{G}_{*}}+\frac{2}{G
					_3+\bar{G}_{*}}\right)\right)^{-1}\!\!-\!\bar{G}_{*}-\bar{G}=0$} \\
			&&\\
			 & {\small{$\bar{K}_{*}=4\bar{G}$}} & {\small{$\bar{G}_{*}=\bar{G}\frac{9\bar{K}+8\bar{G}}{6(\bar{K}+2\bar{G})}$}}       \\
			\hline
		\end{tabular}\\[.051in]
	}
\end{table} 

\section{Detailed results of atomistic simulations}\label{App} 

In this Appendix detailed results of atomistic simulations for eight samples of six metals of hcp symmetry are collected is subsequent subsections. For each metal the following convention is used,
\begin{eqnarray}
\centering
\left[C_{KL}\right]=\left[
\begin{array}{cccccc}
{C_{1111}} & {C_{1122}} & {C_{1133}} & {C_{1123}} & {C_{1131}} & {C_{1112}} \\
& {C_{2222}} & {C_{2233}} & {C_{2223}} & {C_{2231}} & {C_{2212}} \\
&  & {C_{3333}} & {C_{3323}} & {C_{3331}} & {C_{3312}} \\
& & & {C_{2323}} & {C_{2331}} & {C_{2312}} \\
\multicolumn{4}{c}{\textbf{Sym.}} & {C_{3131}} & {C_{3112}} \\
& & & & & {C_{1212}} \\
\end{array}
\right]\,.
\label{eqn:CuCij}
\end{eqnarray} 
The quantitative data describing analysed samples are collected in the first table, while the calculated 21 components of the anisotropic elasticity tensor for each sample (the Voigt notation (\ref{eqn:CuCij}) is used) are given in the second table.

{

\subsection{Nanocrystalline ruten}
   
	\begin{table}[H] 
	\caption{Ruten: Volume (\AA$^3$), box lengths: a,b,c (\AA), number of atoms, average grain diameter $d$ (\AA), fraction of transient shell atoms $f_{0}$ (\ref{def:fsa}), average cohesive energy $E_{c}$\,(eV/atom) of analysed computational samples.  Used EAM potential \cite{Fortini2008} with \emph{cutoff radius}=6.5(\AA).}
	\label{tab:SamplesRu}
	\centering
	\renewcommand{\arraystretch}{1.5}
	\tiny 
	\begin{tabular}{|c c c c c c c c c|}
		\hline Sample & V & a & b & c & No.of atoms & $d$ & $f_{0}$ &  E$_{c}$\\ 
		\hline Monocrystal & 44.56 & 2.70 & 4.68 & 4.29 & 4 & &  & -6.86 \\
		S-128-BCC & 143764.4 & 52.39 & 52.36 & 52.41 & 10324 & 12.90 & 1.00 &-6.60 \\
		M-128-BCC & 6841505.5 & 189.85 & 189.81 & 189.85 & 496965 & 46.74 & 0.59 &-6.75 \\
		M-16-BCC & 6824700.3 & 189.63 & 189.73 & 189.69 & 497225 & 93.39 & 0.34 & -6.80 \\
		M-54-BCC & 6839237.3 & 189.81 & 189.85 & 189.79 & 497232 & 62.31 & 0.47 &-6.77 \\
		M-250-BCC & 6877358.3 & 190.13 & 190.18 & 190.19 & 497416 & 37.45 & 0.69 &-6.72 \\
		M-125-Random & 6867930 & 190.11 & 190.10 & 190.04 & 497109 & 47.17 & 0.59 &-6.74 \\
		L-16-BCC & 54324813 & 378.74 & 378.74 & 378.72 & 3976847 & 186.48 & 0.18 & -6.83 \\
		\hline 
	\end{tabular}
\end{table}

\begin{table}[H] 
	\caption{{{Ruten: Elasticity tensors $\bar{\mathbb{C}}$\,[GPa]  of analysed samples (for notation used see Eq. (\ref{eqn:CuCij}))}}.}
	\label{tab:Cij-cRu}
	\begin{threeparttable}[b]
	\centering
	\tiny

	\begin{tabular}{ c c }
		\hline
		{Monocrystal} & {small sample, 128 grains in BCC system, 10324 atoms} \\
		& {(S-128-BCC)}	  \\ \hline
		$\begin{bmatrix}  546.54 & 169.87 & 170.852 & 0       &       0 & 0 \\
		                         & 546.54 & 170.85  & 0       &       0 & 0 \\
	                             &        & 619.068 & 0       & 0       & 0 \\
		                         &        &         & 199.58  &       0 & 0 \\
		\multicolumn{4}{c}{\textbf{Sym.}}                     & 199.58  & 0 \\
		                         &        &         &         &         & 188.34 \end{bmatrix}$ &	
		
		$\begin{bmatrix}  260.98 & 106.61  & 102.56  & 2.76    & 1.52    & 0.26 \\
		                         & 268.14  & 101.94  & 4.56    & -2.03   & -0.60 \\
		                         &         & 265.35  & 0.21    & -1.95   & -0.41 \\
		                         &         &         & 80.207  & -0.38   & -2.00 \\
		\multicolumn{4}{c}{\textbf{Sym.}}                      & 79.738  & 0.33 \\
		                         &         &         &         &         & 81.049 \end{bmatrix}$\\  		
		\hline {medium sample, 128 grains in BCC system, 496965 atoms} & {medium sample, 16 grains in BCC system,  497225 atoms} \\  
		{(M-128-BCC)} & {(M-16-BCC)} \\  \hline
		$\begin{bmatrix}  367.49 & 154.52  & 171.95 & 11.07   & -11.72  & 5.32 \\
		                         & 356.65  & 159.56 & 13.67   & 10.03   & 3.22 \\
		                         &         & 375.60 & -1.05   & -0.57   & -7.54 \\
		                         &         &        & 109.53  & 10.96   & -3.50 \\
		\multicolumn{4}{c}{\textbf{Sym.}}                     & 90.75   & 8.25 \\
		                         &         &        &         &         & 99.73 \end{bmatrix}$ &
		
		$\begin{bmatrix}  391.47 & 134.67  & 133.33 & 1.98    & -3.09   & -3.98 \\
		                         & 395.65  & 138.81 & 3.85    & -4.21   & -0.74 \\
		                         &         & 393.52 & 2.43    & -6.72   & -2.61 \\
		                         &         &        & 138.01  & -4.42   & -4.58 \\
		\multicolumn{4}{c}{\textbf{Sym.}}                     & 133.14  & -0.76 \\
		                         &         &        &         &         & 143.99 \end{bmatrix}$\\		
		
		\hline {medium sample, 54 grains in BCC system, 497232 atoms} & {medium sample, 250 grains in BCC system, 497416 atoms} \\
		{(M-54-BCC)} & {(M-250-BCC)} \\ \hline
		$\begin{bmatrix}  325.82 & 148.29 & 149.72 & -5.69 & -4.08 & 1.70 \\
	                             & 360.59 & 158.60 & -11.83& 0.51  & 2.05 \\
		                         &        & 368.67 & -5.13 & 6.94  & -9.46 \\
		                         &        &        & 129.22& -5.70 & -5.24 \\
		\multicolumn{4}{c}{\textbf{Sym.}}                  & 125.47& 0.56 \\
		                         &        &        &       &       & 121.66 \end{bmatrix}$ &
		
		$\begin{bmatrix}  301.66 & 137.14 & 126.93 & -13.11 & 3.92 & -5.82 \\
		                         & 271.96 & 120.63 & 7.93   & 4.64 & -7.84 \\
		                         &        & 300.11 & -0.78  & 10.04& 4.38 \\
		                         &        &        & 86.92  & 11.25& 3.96 \\
		\multicolumn{4}{c}{\textbf{Sym.}}                   & 88.70& 7.86 \\
	                             &        &        &        &      & 104.82 \end{bmatrix}$\\
		
		\hline {medium sample, 125 grains in random system, 497109 atoms} & {large sample, 16 grains in BCC system, 3976847 atoms} \\
		{(M-125-Random)} & {(L-16-BCC)} \\ \hline
		$\begin{bmatrix}  322.38 & 110.49 & 101.51 & 4.24 & 1.33 & -9.67 \\
		                         & 308.70 & 127.41 & 2.97 & 6.51 & -17.32 \\
		                         &        & 328.57 & 7.05 & -8.72& -5.17 \\
		                         &        &        & 92.16& 7.38 & 2.93 \\
		\multicolumn{4}{c}{\textbf{Sym.}}                 &101.45& -1.57 \\
		                         &        &        &      &      & 86.768 \end{bmatrix}$ &
		
		$\begin{bmatrix} 456.11 & 155.18 & 149.84 & -0.09  & -0.97 & -3.25 \\
		                        & 450.08 & 151.72 & 1.57   & 1.78  & -2.35 \\
		                        &        &  442.07& 1.43   & -0.74 & -1.75 \\
		                        &        &        & 160.56 & -0.79 & 0.99 \\
		\multicolumn{4}{c}{\textbf{Sym.}}                  & 156.78& 0.43 \\
	                         	&        &        &         &      & 162.96 \end{bmatrix}$\\
		
		\hline
	\end{tabular}
\end{threeparttable}
\end{table}

\subsection{Nanocrystalline titanium}

\begin{table}[H] 
	\caption{Titanium: Volume (\AA$^3$), box lengths: a,b,c (\AA), number of atoms, average grain diameter $d$ (\AA), fraction of transient shell atoms $f_{0}$ (\ref{def:fsa}), average cohesive energy $E_{c}$\,(eV/atom) of analysed computational samples.}
	\label{tab:SamplesTi}
	\centering
	\renewcommand{\arraystretch}{1.5}
	\tiny 
	\begin{tabular}{|c c c c c c c c c|}
		\hline Sample & V & a & b & c & No.of atoms & $d$ & $f_{0}$ &  E$_{c}$\\ 
		\hline Monocrystal & 70.69 & 2.95 & 5.11 & 4.68 & 4 & &  & -4.85 \\
		small-128-BCC & 194855.95 & 58.06 & 58.01 & 57.85 & 11024 & 14.27 & 1.00 &-4.74 \\
		M-128-BCC & 8782331.5  & 206.34 & 206.35 & 206.26 & 494687 & 50.79 & 0.55 &-4.79 \\
		M-16-BCC & 8768351  & 206.16 & 206.21 & 206.25 & 494702 & 101.53 & 0.31 & -4.82 \\
		M-54-BCC & 8779701.6 & 206.28 & 206.35 & 206.25 & 494820 & 67.72 & 0.44 &-4.80 \\
		M-250-BCC & 8786488 & 206.34 & 206.29 & 206.42 & 494813 & 40.64 & 0.65 &-4.78 \\
		M-125-Random & 8781498.7 & 206.38 & 206.28 & 206.27 & 494670 & 51.19 & 0.55 &-4.79 \\
		L-16-BCC & 70048712 & 412.25 & 412.16 & 412.26 & 3957154 & 202.97 & 0.17 & -4.83 \\
		\hline 
	\end{tabular}
\end{table} 

\begin{table}[H] 
	\caption{{{Titanium: Elasticity tensors $\bar{\mathbb{C}}$\,[GPa]  of analysed samples (for notation used see Eq. (\ref{eqn:CuCij}))}}.}
	\label{tab:Cij-cTi}
	\begin{threeparttable}[b]
	\centering
	\tiny
	\begin{tabular}{ c c }
		\hline
		{Monocrystal} & {small sample, 128 grains in BCC system, 11024 atoms} \\
			& {(S-128-BCC)}	  \\ \hline
		$\begin{bmatrix}  {171.46} & {84.23}  & {77.07}  & 0       &       0 & 0 \\
		                           & {171.46} & {77.07}  & 0       &       0 & 0 \\
		                           &          & {189.96} & 0       & 0       & 0 \\
		                           &          &          & {52.79} &       0 & 0 \\
		\multicolumn{4}{c}{\textbf{Sym.}}                          & {52.79} & 0 \\
	                               &          &          &         &         & {43.62} \end{bmatrix}$ &	
		
		$\begin{bmatrix}  {123.29} & {78.92}  & {79.86}  & {-0.56} & {-0.59} & {-0.52} \\
		                           & {121.30} & {79.23}  & {0.03}  & {-0.44} & {0.23} \\
	                               &          & {122.54} & {-0.13} & {-0.41} & {-0.37} \\
		                           &          &          & {20.89} & {-0.56} & {0.34} \\
		\multicolumn{4}{c}{\textbf{Sym.}}                          & {21.42} & {0.01} \\
		                           &          &          &         &         & {22.01} \end{bmatrix}$\\  
		
		\hline {medium sample, 128 grains in BCC system, 494687 atoms} & {medium sample, 16 grains in BCC system,  494702 atoms} \\  
		{(M-128-BCC)} & {(M-16-BCC)} \\  \hline
		$\begin{bmatrix}  {142.10} & {79.56}  & {83.57} & {-3.10} & {2.22}  & {5.97} \\
	                               & {135.40} & {85.37} & {0.59}  & {0.37}  & {1.80} \\
		                           &          & {132.26}& {1.58}  & {-1.41} & {-0.81} \\
		                           &          &         & {26.24} & {2.23}  & {0.67} \\
		\multicolumn{4}{c}{\textbf{Sym.}}                         & {28.46} & {-1.64} \\
	                               &          &         &         &         & {36.56} \end{bmatrix}$ &
		
		$\begin{bmatrix}  {152.64} & {80.87}  & {80.26} & {0.47}  & {0.80}  & {-0.68} \\
		                           & {149.80} & {80.34} & {-0.67} & {-1.23} & {1.22} \\
		                           &          & {148.97}& {0.52}  & {-0.46} & {1.10} \\
		                           &          &         & {38.13} & {-2.30} & {-2.23} \\
		\multicolumn{4}{c}{\textbf{Sym.}}                         & {36.91} & {1.46} \\
	                               &          &          &        &         & {34.19} \end{bmatrix}$\\		
		
		\hline {medium sample, 54 grains in BCC system, 494820 atoms} & {medium sample, 250 grains in BCC system, 494813 atoms} \\
          {(M-54-BCC)} & {(M-250-BCC)} \\ \hline
		$\begin{bmatrix}  {135.42} & {84.15} & {76.08} & {-5.73} & {4.01} & {-0.56} \\
		                           & {138.70} & {85.13} & {1.70} & {-2.45} & {-1.57} \\
		                           &          & {139.64} & {-3.24}  & {4.10}& {-0.40} \\
		                           &          &          & {25.53} & {5.35} & {-0.34} \\
		\multicolumn{4}{c}{\textbf{Sym.}}                          & {29.27}& {0.79} \\
		                           &          &          &         &        & {33.88} \end{bmatrix}$ &
		
		$\begin{bmatrix}  {123.96} & {79.66} & {80.19} & {1.73} & {1.62} & {0.80} \\
		                           & {134.26} & {79.16} & {3.12} & {1.19} & {-3.23} \\
	                               &          & {130.74} & {-3.20}  & {0.38}& {1.96} \\
	             	               &          &          & {25.26} & {0.50} & {6.38} \\
		\multicolumn{4}{c}{\textbf{Sym.}}                          & {23.85}& {0.52} \\
		                           &          &          &         &        & {20.587} \end{bmatrix}$\\
		
		\hline {medium sample, 125 grains in random system, 494670 atoms} & {large sample, 16 grains in BCC system, 3957154 atoms} \\
		{(M-125-Random)} & {(L-16-BCC)} \\ \hline
		$\begin{bmatrix}  {137.88} & {80.55} & {78.25} & {-2.01} & {-1.99} & {-2.31} \\
		                           & {133.908} & {80.32} & {1.90} & {3.78} & {-0.83} \\
		                           &          & {138.98} & {0.25}  & {0.72}& {0.89} \\
		                           &          &          & {27.08} & {-1.38}& {-1.35} \\
		\multicolumn{4}{c}{\textbf{Sym.}}                         & {28.74}& {0.03} \\
		                           &          &          &         &        & {28.48} \end{bmatrix}$ &
		
		$\begin{bmatrix} {161.47} & {79.46} & {80.93} & {0.39}  & {-0.78}  & {-0.61} \\
		                          & {162.70} & {80.79} & {0.76} & {0.35} & {-0.23} \\
		                          &          &  {159.74}& {-0.90}  & {-0.32}  & {-0.10} \\
		                          &          &          & {42.17} & {-0.55}  & {-0.56} \\
		\multicolumn{4}{c}{\textbf{Sym.}}                         & {41.57} & {0.00} \\
		                          &          &          &         &         & {42.10} \end{bmatrix}$\\
		
		\hline
	\end{tabular}
\end{threeparttable}
\end{table}

\subsection{Nanocrystalline cobalt}

\begin{table}[H] 
	\caption{Cobalt: Volume (\AA$^3$), box lengths: a,b,c (\AA), number of atoms, average grain diameter $d$ (\AA), fraction of transient shell atoms $f_{0}$ (\ref{def:fsa}), average cohesive energy $E_{c}$\,(eV/atom) of analysed computational samples.  Used EAM potential \cite{Pun2012} with \emph{cutoff radius}=6.5 (\AA).}
	\label{tab:SamplesCo}
	\centering
	\renewcommand{\arraystretch}{1.5}
	\tiny 
		\begin{tabular}{|c c c c c c c c c|}
		\hline Sample & V & a & b & c & No.of atoms & $d$ & $f_{0}$ &  E$_{c}$\\ 
		\hline Monocrystal & 44.56 & 2.52 & 4.36 & 4.06 & 4 & &  & -4.39 \\
		small-128-BCC & 147622.4 & 52.74 & 52.84 & 52.97 & 12692 & 13.01 & 1.00 &-4.27 \\
		M-128-BCC & 5775045 & 179.44 & 179.41 & 179.39 & 506233 & 44.17 & 0.61  &-4.33 \\
		M-16-BCC & 5716172.5 & 178.82 & 178.78 & 178.81 & 506408 & 88.04 & 0.36 & -4.36 \\
		M-54-BCC & 5745458.3 & 179.14 & 179.04 & 179.14 & 506071 & 58.79 & 0.50 &-4.34 \\
		M-250-BCC & 5801318.1 & 179.75 & 179.59 & 179.72 & 506367 & 35.39 & 0.71 &-4.32 \\
		M-125-Random & 5779664.5 & 179.49 & 179.42 & 179.46 & 506180 & 44.53 & 0.61 &-4.33 \\
		L-16-BCC & 45430931 & 356.78 & 356.81 & 356.87 & 4049913 & 175.69 & 0.19 & -4.37 \\
		\hline 
	\end{tabular}
	\end{table} 

\begin{table}[H] 
	\caption{{{Cobalt: Elasticity tensors $\bar{\mathbb{C}}$\,[GPa]  of analysed samples (for notation used see Eq. (\ref{eqn:CuCij}))}}.}
	\label{tab:Cij-cCo}
	\begin{threeparttable}[b]
	\centering
	\tiny
	\begin{tabular}{ c c }
		\hline
		{Monocrystal} & {small sample, 128 grains in BCC system, 12692 atoms} \\
		& {(S-128-BCC)}	  \\ \hline
		$\begin{bmatrix}  310.01 & 145.67 & 119.48  & 0       &       0 & 0 \\
		                         & 310.01 & 119.48  & 0       &       0 & 0 \\
		                         &        & 357.51  & 0       & 0       & 0 \\
		                         &        &         & 92.54   &       0 & 0 \\
		\multicolumn{4}{c}{\textbf{Sym.}}                     & 92.54   & 0 \\
		                         &        &          &        &         & 82.17 \end{bmatrix}$ &	
		
		$\begin{bmatrix}  243.84 & 174.23 & 173.08 & -1.25   & 1.06    & -0.74 \\
		                         & 240.30 & 175.27 & 1.33    & -1.39   & 0.54 \\
		                         &        & 242.47 & -0.70   & 0.66    & 0.40 \\
		                         &        &        & 35.80   & 1.08    & 0.47 \\
		\multicolumn{4}{c}{\textbf{Sym.}}                    & 36.07   & 0.58 \\
		                         &        &        &         &         & 34.62 \end{bmatrix}$\\  
		
		\hline {medium sample, 128 grains in BCC system, 506233 atoms} & {medium sample, 16 grains in BCC system, 506408 atoms} \\  
		{(M-128-BCC)} & {(M-16-BCC)} \\  \hline
		$\begin{bmatrix}  264.68 & 163.80  & 162.70 & 3.63  & 0.42  & 0.12 \\
		                         & 268.47  & 163.30 & 1.70  & 1.38  & 2.65 \\
		                         &         & 265.43 & -1.04 & 1.04  & 0.38 \\
	                        	 &         &        & 58.85 & -0.18 & 0.83 \\
		\multicolumn{4}{c}{\textbf{Sym.}}                   & 56.66 & -0.30 \\
		                         &         &        &       &       & 57.70 \end{bmatrix}$ &
		
		$\begin{bmatrix}  280.92 & 150.26  & 156.34 & 0.03    & -2.66   & -2.48 \\
		                         & 278.34  & 157.49 & -1.48   & 1.07    & -4.16 \\
		                         &         & 273.10 & 1.65    & 0.96    & 5.33 \\
		                         &         &        & 68.97   & -1.27   & 0.39 \\
		\multicolumn{4}{c}{\textbf{Sym.}}                     & 68.88   & -1.93 \\
	               	             &         &        &         &         & 68.66 \end{bmatrix}$\\		
		
		\hline {medium sample, 54 grains in BCC system, 506071 atoms} & {medium sample, 250 grains in BCC system, 506367 atoms} \\
		{(M-54-BCC)} & {(M-250-BCC)} \\ \hline
		$\begin{bmatrix}  271.98 & 157.78 & 159.71 & -2.83   & 0.96  & -1.46 \\
		                         & 271.10 & 159.86 & -2.45   & 1.46  & 1.24 \\
		                         &        & 272.01 & 0.75    & -0.85 & -0.06 \\
		                         &        &        & 63.139  & 1.11  & 0.45 \\
		\multicolumn{4}{c}{\textbf{Sym.}}                    & 63.48 & -0.04 \\
		                         &        &        &         &       & 61.88 \end{bmatrix}$ &
		
		$\begin{bmatrix}  260.16 & 167.37 & 166.43  & 0.18   & -2.13  & 1.06 \\
		                         & 251.61 & 164.01  & 2.30   & 2.01   & 0.22 \\
		                         &        & 260.67  & -0.80  & 1.88   & -0.99 \\
		                         &        &         & 51.01  & 0.04   & 0.88 \\
		\multicolumn{4}{c}{\textbf{Sym.}}                    & 51.58 & 1.53 \\
		                         &        &         &        &        & 52.48 \end{bmatrix}$\\
		
		\hline {medium sample, 125 grains in random system, 506180 atoms} & {large sample, 16 grains in BCC system, 4049913 atoms} \\
		{(M-125-Random)} & {(L-16-BCC)} \\ \hline
		$\begin{bmatrix}  268.50 & 165.27 & 160.30   & -0.85  & 1.54  & 5.79 \\
		                         & 271.05 & 156.73   & -2.98  & -3.17 & -3.76 \\
		                         &          & 271.88 & 2.09   & -1.64 & -1.96 \\
		                         &          &        & 47.918 & -0.03 & -2.21 \\
		\multicolumn{4}{c}{\textbf{Sym.}}                     & 51.42 & -5.74 \\
		                         &          &        &        &       & 37.45 \end{bmatrix}$ &
		
		$\begin{bmatrix} 292.95 & 143.53 & 149.04 & -0.77  & -1.32  & -0.60 \\
		                        & 295.14 & 147.50 & 1.85   & 0.83   & -1.96 \\
		                        &        &  287.36& -0.41  & 0.13   & 1.21 \\
		                        &        &        & 74.48  & -1.35  & -0.50 \\
		\multicolumn{4}{c}{\textbf{Sym.}}                  & 77.50  & -0.38 \\
		                        &        &        &        &        & 79.39 \end{bmatrix}$\\
		\hline
	\end{tabular}
\end{threeparttable}
\end{table}

\subsection{Nanocrystalline zirconium}

\begin{table}[H] 
	\caption{Zirconium: Volume (\AA$^3$), box lengths: a,b,c (\AA), number of atoms, average grain diameter $d$ (\AA), fraction of transient shell atoms $f_{0}$ (\ref{def:fsa}), average cohesive energy $E_{c}$\,(eV/atom) of analysed computational samples. Used EAM potential \cite{Mendelev2007} with \emph{cutoff radius}=7.6 (\AA).}
	\label{tab:SamplesZr}
	\centering
	\renewcommand{\arraystretch}{1.5}
	\tiny 
	\begin{tabular}{|c c c c c c c c c|}
		\hline Sample & V & a & b & c & No.of atoms & $d$ & $f_{0}$ &  E$_{c}$\\ 
		\hline Monocrystal & 93.75 & 3.23 & 5.59 & 5.19 & 4 & &  & -6.02 \\
		S-128-BCC & 264763.83 & 64.23 & 64.11 & 64.29 & 11172 & 15.81 & 0.99 &-5.90 \\
		M-128-BCC & 11630649  & 226.55 & 226.60 & 226.55 & 492603 & 55.78 & 0.52 &-5.95 \\
		M-16-BCC & 11592882  & 226.34 & 226.32 & 226.31 & 492442 & 111.44 & 0.29 & -5.98 \\
		M-54-BCC & 11611827 & 226.35 & 226.42 & 226.57 & 492413 & 74.33 & 0.41 &-5.97 \\
		M-250-BCC & 11641724 & 226.64 & 226.64 & 226.64 & 492635 & 44.64 & 0.61 &-5.94 \\
		M-125-Random & 11626043 & 226.59 & 226.51 & 226.52 & 492598 & 56.21 & 0.51 &-5.95 \\
		L-16-BCC & 92583204 & 452.41 & 452.34 & 452.41 & 3940813 & 222.74 & 0.15 & -6.00 \\
		\hline 
	\end{tabular}
\end{table}

\begin{table}[H] 
	\caption{{{Zirconium: Elasticity tensors $\bar{\mathbb{C}}$\,[GPa]  of analysed samples (for notation used see Eq. (\ref{eqn:CuCij}))}}.}
	\label{tab:Cij-cZr}
	\begin{threeparttable}[b]
	\centering
	\tiny
	\begin{tabular}{ c c }
		\hline
		{Monocrystal} & {small sample, 128 grains in BCC system, 11172 atoms} \\
		& {(S-128-BCC)}	  \\ \hline
		$\begin{bmatrix}  174.26 & 109.69 & 80.54  & 0       &       0 & 0 \\
		                         & 174.26 & 80.54  & 0       &       0 & 0 \\
		                         &        & 211.40 & 0       & 0       & 0 \\
		                         &        &        & 46.45   &       0 & 0 \\
		\multicolumn{4}{c}{\textbf{Sym.}}                    & 46.45   & 0 \\
		                         &        &        &         &         & 32.29 \end{bmatrix}$ &	
		
		$\begin{bmatrix}  108.52 & 74.07  & 73.50  & 0.36    & 0.93    & 0.29 \\
		                         & 108.11 & 74.37  & -0.56   & 0.05    & 0.69 \\
		                         &        & 109.46 & 0.47    & -1.07   & -1.09 \\
		                         &        &        & 16.87   & 0.30    & 0.06 \\
		\multicolumn{4}{c}{\textbf{Sym.}}                    & 17.62   & -0.17 \\
		                         &        &        &         &         & 17.66 \end{bmatrix}$\\  
		
		\hline {medium sample, 128 grains in BCC system, 492603 atoms} & {medium sample, 16 grains in BCC system, 492442 atoms} \\  
		{(M-128-BCC)} & {(M-16-BCC)} \\  \hline
		$\begin{bmatrix}  125.81 & 85.97   & 81.95  & -1.78 & -1.28 & 3.01 \\
		                         & 130.15  & 85.26  & 0.28  & -0.09 & 2.51 \\
		                         &         & 129.99 & 1.93  & -0.67 & -2.85 \\
		                         &         &        & 22.31 & -0.68 & -0.29 \\
		\multicolumn{4}{c}{\textbf{Sym.}}                   & 24.27 & -0.14 \\
		                         &         &        &       &       & 26.27 \end{bmatrix}$ &
		
		$\begin{bmatrix}  158.31 & 85.18  & 86.49 & 1.19    & -1.62   & -2.28 \\
		                         & 152.80 & 90.89 & -0.66   & 0.20    & -0.64 \\
		                         &        & 141.00& -6.58   & 0.65    & 2.74 \\
		                         &        &       & 29.08   & -2.15   & -0.46 \\
		\multicolumn{4}{c}{\textbf{Sym.}}                   & 30.27   & 0.75 \\
		                         &        &       &         &         & 32.56 \end{bmatrix}$\\		
		
		\hline {medium sample, 54 grains in BCC system, 492413 atoms} & {medium sample, 250 grains in BCC system, 492635 atoms} \\
		{(M-54-BCC)} & {(M-250-BCC)} \\ \hline
		$\begin{bmatrix}  139.04 & 87.52  & 87.94  & 0.69    & 0.88  & -0.65 \\
		                         & 138.54 & 84.56  & 0.26    & -0.46 & -2.72 \\
		                         &        & 134.44 & -1.52   & -5.56 & 0.46 \\
		                         &        &        & 27.44   & -2.35 & -1.57 \\
		\multicolumn{4}{c}{\textbf{Sym.}}                    & 26.09 & -2.57 \\
		                         &        &        &         &       & 28.77 \end{bmatrix}$ &
		
		$\begin{bmatrix}  124.18 & 81.47  & 85.98  & -0.61  & -3.39  & -1.19 \\
		                         & 114.63 & 87.33  & 0.99   & 2.22   & 0.13 \\
		                         &        & 118.27 & 0.63   & 1.71   & -3.03 \\
		                         &        &        & 23.61  & 1.09   & -1.20 \\
		\multicolumn{4}{c}{\textbf{Sym.}}                   & 20.59  & 2.97 \\
		                         &        &        &        &        & 13.65 \end{bmatrix}$\\
		
		\hline {medium sample, 125 grains in random system, 492598 atoms} & {large sample, 16 grains in BCC system, 3940813 atoms} \\
		{(M-125-Random)} & {(L-16-BCC)} \\ \hline
		$\begin{bmatrix}  132.24 & 82.33  & 83.51    & -0.23  & 0.58  & 1.27 \\
		                         & 132.48 & 80.28    & -1.14  & 1.49  & 0.63 \\
		                         &        & 134.34   & 0.31   & -0.45 & 0.20 \\
		                         &        &          & 20.12  & -0.21 & -2.13 \\
		\multicolumn{4}{c}{\textbf{Sym.}}                     & 23.98 & 0.67 \\
		                         &        &          &        &       & 24.53 \end{bmatrix}$ &
		
		$\begin{bmatrix} 160.54 & 90.40  & 93.31  & 0.00   & -0.69  & -0.73 \\
		                        & 162.52 & 92.95  & 0.20   & -0.01  & -1.78 \\
		                        &        &  157.70& -0.01  & 0.98   & 1.26 \\
		                        &        &        & 34.62  & -1.91  & -0.36 \\
		\multicolumn{4}{c}{\textbf{Sym.}}                  & 35.17  & 0.13 \\
		                        &        &        &        &        & 37.60 \end{bmatrix}$\\
		\hline
	\end{tabular}
\end{threeparttable}
\end{table}

\subsection{Nanocrystalline magnesium}

\begin{table}[H] 
	\caption{Magnesium: Volume (\AA$^3$), box lengths: a,b,c (\AA), number of atoms, average grain diameter $d$ (\AA), fraction of transient shell atoms $f_{0}$ (\ref{def:fsa}), average cohesive energy $E_{c}$\,(eV/atom) of analysed computational samples. Used EAM potential \cite{Johnson2004} with \emph{cutoff radius}=7.15 (\AA).}
	\label{tab:SamplesMg}
	\centering
	\renewcommand{\arraystretch}{1.5}
	\tiny 
	\begin{tabular}{|c c c c c c c c c|}
		\hline Sample & V & a & b & c & No.of atoms & $d$ & $f_{0}$ &  E$_{c}$\\ 
		\hline Monocrystal & 92.37 & 3.20 & 5.54 & 5.21 & 4 & &  & -1.55 \\
		small-128-BCC & 243204.72 & 62.38 & 62.43 & 62.46 & 10188 & 15.37 & 0.99 & -1.51 \\
		M-128-BCC & 11683248 & 226.94 & 226.87 & 226.92 & 496137 & 55.86 & 0.52 & -1.53 \\
		M-16-BCC & 11588292  & 226.30 & 226.32 & 226.26 & 496294 & 111.42 & 0.29 & -1.54 \\
		M-54-BCC & 11644640 & 226.63 & 226.69 & 226.66 & 496382 & 74.40 & 0.41	 & -1.53 \\
		M-250-BCC & 11388363 & 224.96 & 225.02 & 224.98 & 496234 & 44.31 & 0.61 & -1.49 \\
		M-125-Random & 11387887 & 224.98 & 224.96 & 225.00 & 496477 & 55.83 & 0.52 & -1.50 \\
		L-16-BCC & 92228298 & 451.88 & 451.82 & 451.73 & 3970217 & 222.46 & 0.15 & -1.54 \\
		\hline 
	\end{tabular}
\end{table}

\begin{table}[H] 
	\caption{{{Magnesium: Elasticity tensors $\bar{\mathbb{C}}$\,[GPa]  of analysed samples (for notation used see Eq. (\ref{eqn:CuCij}))}}.}
	\label{tab:Cij-cMg}
	\begin{threeparttable}[b]
		\centering
		\tiny
		\begin{tabular}{ c c }
			\hline
			{Monocrystal} & {small sample, 128 grains in BCC system, 10188 atoms} \\
			& {(S-128-BCC)}	  \\ \hline
			$\begin{bmatrix}  55.88 & 28.70 & 20.19  & 0       &       0 & 0 \\
			& 55.88 & 20.19  & 0       &       0 & 0 \\
			&       & 69.40  & 0       & 0       & 0 \\
			&       &        & 13.86   &       0 & 0 \\
			\multicolumn{4}{c}{\textbf{Sym.}}                  & 13.86   & 0 \\
			&       &        &         &         & 13.59 \end{bmatrix}$ &	
			
			$\begin{bmatrix}  41.98 & 28.92  & 28.57  & -0.36   & -0.18   & -0.31 \\
			& 41.60  & 28.78  & -0.02   & -0.62   & 0.04 \\
			&        & 42.27  & -0.53   & -0.10   & 0.06 \\
			&        &        & 6.13    & -0.36   & -0.60 \\
			\multicolumn{4}{c}{\textbf{Sym.}}                   & 5.68    & -0.05 \\
			&        &        &         &         & 6.62 \end{bmatrix}$\\  
			
			\hline {medium sample, 128 grains in BCC system, 496137 atoms} & {medium sample, 16 grains in BCC system, 496294 atoms} \\  
			{(M-128-BCC)} & {(M-16-BCC)} \\  \hline
			$\begin{bmatrix}  44.51 & 28.38   & 27.90  & 1.33  & 0.13  & 0.11 \\
			& 46.38   & 27.37  & 0.12  & 0.38  & 0.26 \\
			&         & 46.79  & -0.11 & -0.12 & -0.36 \\
			&         &        & 9.05  & -0.05 & 0.33 \\
			\multicolumn{4}{c}{\textbf{Sym.}}                  & 9.43  & 0.41 \\
			&         &        &       &       & 9.83 \end{bmatrix}$ &
			
			$\begin{bmatrix}  50.12 & 26.20  & 27.33 & -0.23   & -0.26   & 0.08 \\
			& 50.62  & 26.74 & 0.62    & 0.06    & -0.15 \\
			&        & 49.20 & -0.36   & 0.04    & -0.17 \\
			&        &       & 11.42   & -0.61   & -0.08 \\
			\multicolumn{4}{c}{\textbf{Sym.}}                  & 12.06   & -0.13 \\
			&        &       &         &         & 11.80 \end{bmatrix}$\\		
			
			\hline {medium sample, 54 grains in BCC system, 496382 atoms} & {medium sample, 250 grains in BCC system, 496234 atoms} \\
			{(M-54-BCC)} & {(M-250-BCC)} \\ \hline
			$\begin{bmatrix}  48.55 & 26.57  & 27.32  & -0.05   & -0.47 & -0.33 \\
			& 48.23  & 27.92  & 0.27    & 0.43  & 0.22 \\
			&        & 46.90  & -0.20   & -0.56 & 0.46 \\
			&        &        & 10.33   & -0.18 & -0.12 \\
			\multicolumn{4}{c}{\textbf{Sym.}}                   & 10.25 & 0.56 \\
			&        &        &         &       & 9.22 \end{bmatrix}$ &
			
			$\begin{bmatrix}  44.78 & 28.27  & 28.56  & -0.36  & -0.34  & 0.22 \\
			& 44.90  & 27.43  & 0.54   & 0.85   & -0.75 \\
			&        & 44.27  & 0.72   & 0.68   & -1.01 \\
			&        &        & 7.63   & -0.18  & 1.10 \\
			\multicolumn{4}{c}{\textbf{Sym.}}                  & 8.30   & 0.98 \\
			&        &        &        &        & 7.15 \end{bmatrix}$\\
			
			\hline {medium sample, 125 grains in random system, 496477 atoms} & {large sample, 16 grains in BCC system, 3970217 atoms} \\
			{(M-125-Random)} & {(L-16-BCC)} \\ \hline
			$\begin{bmatrix}  46.77   & 26.96  & 27.86 & -0.72  & 0.21  & -0.15 \\
			& 46.76  & 27.89 & 0.28   & -0.13 & -0.24 \\
			&        & 46.53 & 0.57   & -0.44 & -0.08 \\
			&        &       & 8.76   & -0.23 & -0.04 \\
			\multicolumn{4}{c}{\textbf{Sym.}}                   & 9.48  & 0.07 \\
			&        &       &        &       & 9.08 \end{bmatrix}$ &
			
			$\begin{bmatrix} 52.39 & 25.82  & 27.83  & -0.57  & -0.14  & 0.47\\
			& 51.83  & 26.17  & 0.64   & -0.04  & -0.64 \\
			&        &  49.06 & -0.26  & -0.39  & -0.32 \\
			&        &        & 12.37  & -0.49  & 0.12 \\
			\multicolumn{4}{c}{\textbf{Sym.}}                 & 13.04  & -0.25 \\
			&        &        &        &        & 12.68 \end{bmatrix}$\\
		
			\hline
		\end{tabular}
	\end{threeparttable}
\end{table}

\subsection{Nanocrystalline Rhenium}

\begin{table}[H] 
	\caption{Rhenium: Volume (\AA$^3$), box lengths: a,b,c (\AA), number of atoms, average grain diameter $d$ (\AA), fraction of transient shell atoms $f_{0}$ (\ref{def:fsa}), average cohesive energy $E_{c}$\,(eV/atom) of analysed computational samples. Used EAM potential \cite{Setyawan2018} with \emph{cutoff radius}=5.5 (\AA).}
	\label{tab:SamplesRe}
	\centering
	\renewcommand{\arraystretch}{1.5}
	\tiny 
	\begin{tabular}{|c c c c c c c c c|}
		\hline Sample & V & a & b & c & No.of atoms & $d$ & $f_{0}$ &  E$_{c}$\\ 
		\hline Monocrystal & 59.12 & 2.76 & 4.78 & 4.48 & 4 & &  & -8.03 \\
		small-128-BCC & 98039.39 & 46.25 & 46.06 & 46.03 & 6556 & 11.35 & 1.00 & -7.76 \\
		M-128-BCC & 7605402 & 196.68 & 196.63 & 196.66 & 509469 & 48.41 & 0.57 & -7.91 \\
		M-16-BCC & 7663630.7  & 197.10 & 197.19 & 197.18 & 509494 & 97.07 & 0.33 & -7.91 \\
		M-54-BCC & 7592673.7 & 196.53 & 196.51 & 196.60 & 509561 & 64.52 & 0.46 & -7.93 \\
		M-250-BCC & 7617042.6 & 196.74 & 196.79 & 196.74 & 509655 & 38.75 & 0.67 & -7.88 \\
		M-125-Random & 7607934.3 & 196.69 & 196.73 & 196.61 & 509357 & 48.80 & 0.57 & -7.90 \\
		L-16-BCC & 60415965 & 392.37 & 392.39 & 392.41 & 4075464 & 193.20 & 0.176 & -8.00 \\
		\hline 
	\end{tabular}
\end{table}

\begin{table}[H] 
	\caption{{{Rhenium: Elasticity tensors $\bar{\mathbb{C}}$\,[GPa]  of analysed samples (for notation used see Eq. (\ref{eqn:CuCij}))}}.}
	\label{tab:Cij-cRe}
	\begin{threeparttable}[b]
	\centering
	\tiny
	\begin{tabular}{ c c }
		\hline
		{Monocrystal} & {small sample, 128 grains in BCC system, 6556 atoms} \\
		& {(S-128-BCC)}	  \\ \hline
		$\begin{bmatrix}  340.24 & 259.95 & 217.92  & 0       &       0 & 0 \\
		                         & 340.24 & 217.92  & 0       &       0 & 0 \\
		                         &        & 448.68  & 0       & 0       & 0 \\
		                         &        &         & 52.51   &       0 & 0 \\
		\multicolumn{4}{c}{\textbf{Sym.}}                     & 52.51   & 0 \\
		                         &        &         &         &         & 40.14 \end{bmatrix}$ &	
		
		$\begin{bmatrix}  461.71 & 282.82 & 281.18 & -0.51   & 0.42    & -1.85 \\
		                         & 453.84 & 282.99 & -0.87   & 2.46    & 0.83 \\
		                         &        & 460.03 & -1.47   & -0.09   & -0.96 \\
		                         &        &        & 88.35   & -1.12   & 0.04 \\
		\multicolumn{4}{c}{\textbf{Sym.}}                    & 88.80   & 0.17 \\
		                         &        &        &         &         & 87.31 \end{bmatrix}$\\  
		
		\hline {medium sample, 128 grains in BCC system, 509469 atoms} & {medium sample, 16 grains in BCC system, 509494 atoms} \\  
		{(M-128-BCC)} & {(M-16-BCC)} \\  \hline
		$\begin{bmatrix}  382.27 & 250.57  & 251.55  & 0.01  & -0.11  & -0.74 \\
		                         & 386.13  & 260.17  & 3.67  & -0.71  & -0.48 \\
		                         &         & 380.78  & 0.78  & -6.04  & 0.36 \\
		                         &         &         & 66.37 & -1.21  & 0.67 \\
		\multicolumn{4}{c}{\textbf{Sym.}}                    & 69.99 & -2.07 \\
		                         &         &         &       &       & 65.36 \end{bmatrix}$ &
		
		$\begin{bmatrix}  323.32 & 226.52  & 236.74 & -1.26   & -2.20   & 2.87 \\
		                         & 303.72  & 218.24 & -13.17  & -3.32   & 2.32 \\
		                         &         & 303.63 & -2.09   & -10.00  & -1.46 \\
		                         &         &        & 46.94   & 1.11    & 9.81 \\
		\multicolumn{4}{c}{\textbf{Sym.}}                     & 51.21   & -4.98 \\
		                         &         &        &         &         & 63.53 \end{bmatrix}$\\		
		
		\hline {medium sample, 54 grains in BCC system, 509561 atoms} & {medium sample, 250 grains in BCC system, 509655 atoms} \\
		{(M-54-BCC)} & {(M-250-BCC)} \\ \hline
		$\begin{bmatrix}  375.22 & 246.90 & 255.22 & -1.64   & -1.86 & -0.31 \\
		                         & 375.50 & 254.43 & 1.72    & 0.01  & 0.94 \\
		                         &        & 376.65 & -0.41   & -1.83 & 2.00 \\
		                         &        &        & 60.72   & -1.77 & 0.15 \\
		\multicolumn{4}{c}{\textbf{Sym.}}                    & 63.71 & 1.03 \\
		                         &        &        &         &       & 63.73 \end{bmatrix}$ &
		
		$\begin{bmatrix}  377.02 & 261.51  & 257.40 & 0.70   & -10.98 & 5.06 \\
		                         & 376.30  & 257.64 & -4.30  & -2.59  & 1.01 \\
		                         &         & 372.34 & -5.34  & -4.77  & 1.61 \\
		                         &         &        & 68.25  & 3.76   & 2.75 \\
		\multicolumn{4}{c}{\textbf{Sym.}}                    & 64.60  & 2.36 \\
		                         &         &        &        &        & 63.01 \end{bmatrix}$\\
		
		\hline {medium sample, 125 grains in random system, 509357 atoms} & {large sample, 16 grains in BCC system, 4075464 atoms} \\
		{(M-125-Random)} & {(L-16-BCC)} \\ \hline
		$\begin{bmatrix}  393.93 & 257.81 & 255.82 & 2.23   & -1.76 & 0.55 \\
		                         & 392.17 & 253.97 & 2.51   & 0.10  & -1.86 \\
		                         &        & 388.19 & 1.57   & -0.97 & -0.95 \\
		                         &        &        & 67.07  & -0.21 & -1.02 \\
		\multicolumn{4}{c}{\textbf{Sym.}}                   & 66.75 & -0.82 \\
		                         &        &        &        &       & 67.88 \end{bmatrix}$ &
		
		$\begin{bmatrix} 364.23 & 240.56 & 250.59  & -1.46  & -2.04  & -2.05 \\
		                        & 369.84 & 245.76  & 2.45   & 1.43   & -3.95 \\
		                        &        &  352.61 & -0.49  & -1.18  & 0.61 \\
		                        &        &         & 53.53  & -4.16  & -1.18 \\
		\multicolumn{4}{c}{\textbf{Sym.}}                   & 56.49  & -0.86 \\
		                        &        &         &        &        & 57.95 \end{bmatrix}$\\
		\hline
	\end{tabular}
  \end{threeparttable}
\end{table}

}

\section*{References}

\bibliography{References-mm}

\end{document}